\documentclass[traditabstract]{aa}
\usepackage{natbib}
\usepackage{longtable}
\usepackage{graphpap, times,graphicx}
\usepackage{lscape}
\usepackage{ulem}

\bibpunct{(}{)}{;}{a}{}{,}
\newcommand{\be}{\begin{equation}}
\newcommand{\ee}{\end{equation}}

\newcommand{\kms}{km\,s$^{-1}$}
\newcommand{\cog}{{\sc CoG}}
\newcommand{\eleven}{Sk\,-67$^\circ$111}
\newcommand{\new}{LH\,54-425}
\newcommand{\seven}{Sk\,-67$^\circ$107}
\newcommand{\six}{Sk\,-67$^\circ$106}
\newcommand{\four}{Sk\,-67$^\circ$104}
\newcommand{\one}{Sk\,-67$^\circ$101}

\begin{document}
   \title{Interstellar absorptions towards the LMC:}
   \subtitle{Small-scale density variations in Milky Way disc gas}

   \author{S.~Nasoudi-Shoar\inst{1}
          \and P.~Richter\inst{2}
          \and K.S.~de Boer\inst{1}
	  \and B.P.~Wakker\inst{3}
          }

   \offprints{Soroush Nasoudi-Shoar, e-mail{ soroush@astro.uni-bonn.de}}

   \institute{$^{1}$Argelander-Institut f\"ur Astronomie,
Universit\"at Bonn, Auf dem H\"ugel 71, 53121 Bonn, Germany\\
              $^{2}$Institut f\"ur Physik und Astronomie, Universit\"at Potsdam,
Haus 28, Karl-Liebknecht-Str. 24/25, 14476 Golm, Germany\\
	      $^{3}$Department of Astronomy, University of Wisconsin, 475 N Charter St, Madison, WI 53706}

   \date{Accepted 1 June 2010}

\abstract{Observations show that the interstellar medium (ISM) contains sub-structure on scales less than 1\,pc, detected in the form of spatial and temporal variations in column densities or optical depth. Despite the number of detections, the nature and ubiquity of the small-scale structure in the ISM is not yet fully understood. We use UV absorption data mainly from the Far Ultraviolet Spectroscopic Explorer (FUSE) and partly from the Space Telescope Imaging Spectrograph (STIS) of six Large Magellanic Cloud (LMC) stars (\eleven, \new, \seven, \six, \four, and \one) that are all located within 5\arcmin\ of each other, and analyse the physical properties of the Galactic disc gas in front of the LMC on sub-pc scales. 
We analyse absorption lines of a number of ions within the UV spectral range. Most importantly, interstellar molecular hydrogen, neutral oxygen, and fine-structure levels of neutral carbon have been used in order to study changes in the density and the physical properties of the Galactic disc gas over small angular scales. At an assumed distance of 1\,kpc, the 5\arcmin\ separation between \eleven\ and \one\ implies a linear extent of 1.5\,pc. 

We report on column densities of H$_2$, \ion{C}{i}, \ion{N}{i}, O\,{\sc i}, Al\,{\sc ii}, Si\,{\sc ii}, P\,{\sc ii}, S\,{\sc iii}, Ar\,{\sc i}, and \ion{Fe}{ii} in our six lines of sight, as well as \ion{C}{i*}, \ion{C}{i**}, Mg\,{\sc ii}, Si\,{\sc iv}, S\,{\sc ii}, Mn\,{\sc ii}, and Ni\,{\sc ii} for four of them. While most species do not show any significant variation in their column densities, we find an enhancement of almost 2\,dex for H$_2$ from \eleven\ to \one, accompanied by only a small variation in the O\,{\sc i} column density. Based on the formation-dissociation equilibrium, we trace these variations to the actual density variations in the molecular gas.  
On the smallest spatial scale of $<0.08$\,pc, between \seven\ and \new, we find a gas density variation of a factor of $1.8$.  
The line of sight towards \new\ does not follow the relatively smooth change seen from \one\ to \eleven, suggesting that sub-structure might exist on a smaller spatial scale than the linear extent of our sight-lines.

The results show that we sample a mix of both neutral and ionised gas in our six lines of sight. Towards \one\ to \seven, we derive the temperature $T_{\rm exc}\simeq 70$\,K for the inner self-shielded part of the gas based on the rotational excitation levels of H$_2$, and an average density of $n_{\rm H}\simeq 60 \,\rm {cm^{-3}}$, typical of that for CNM. The gas towards \new\ and \eleven\ shows different properties, and $T_{\rm exc}\simeq 200$\,K. 
Our observations suggest that the detected H$_2$ in these six lines of sight (with the extent of $<1.5$\,pc) is not necessarily physically connected, but that we are sampling molecular cloudlets with pathlengths $<0.1-1.8$\,pc and possibly different densities.}
 

\maketitle

\section{Introduction}

\label{sect:Introduction} 

Recent studies of the interstellar medium have shown a number of observed variations on scales smaller than 1\,pc, indicating the existence of small-scale structure. 
These have been reported in the form of temporal variations and differences in column densities over small spatial scales. 
 
Interferometric data of neutral hydrogen in 1976 showed variations in \ion{H}{i} over small scales \citep{Dieter}. Since then a number of studies of small-scale structures in \ion{H}{i} have been performed, by 21-cm absorption lines against pulsars with time variability \citep[e.g.,][]{Clifton88,Frail94}, sampling the gas on scales of tens to hundreds of AUs, or using interferometric observations with the Very Long Baseline Array (VLBA) \citep[e.g.,][]{Faison:Goss} finding optical depth variations on about the same scales. 
 
Optical spectra at high spectral resolution have been taken of stars in globular clusters, providing a grid 
of very close sight-lines (angular scales of only few arcsec) to detect 
structure in interstellar gas on scales down to 0.01\,pc.  
In these studies column density variations have been found in mainly Ca\,{\sc ii} and Na\,{\sc i} lines, and in differential reddening \citep[e.g.][]{Cohen,Bates:1990,Bates:1991,Bates:1995,Meyer:1999,Andrews}. 

Using nearby binaries and multiple star systems with separations $<20$\,arcsec, the smallest projected scales have been probed \citep[e.g.,][]{Meyer:1990,Watson:Meyer,Lauroesch:Meyer2000}. 
 
When measurements on small-scale structure were repeated, temporal variations have been found in several cases.  
Examples have been presented in the above cited investigations by, e.g., \citet{Lauroesch:Meyer2000}, and \citet{Lauroesch:2003}. In some other cases, where a proper motion star was the target, the repeated observations have revealed the spatial variations in the gas on scales smaller than the ones provided by multiple stars and binaries, as the lines of sight sample different parts of the same cloud at different times \citep{Welty2001,Welty2007}.
 
\clearpage
\begin{figure*} 
\centering 
\includegraphics[scale=0.65]{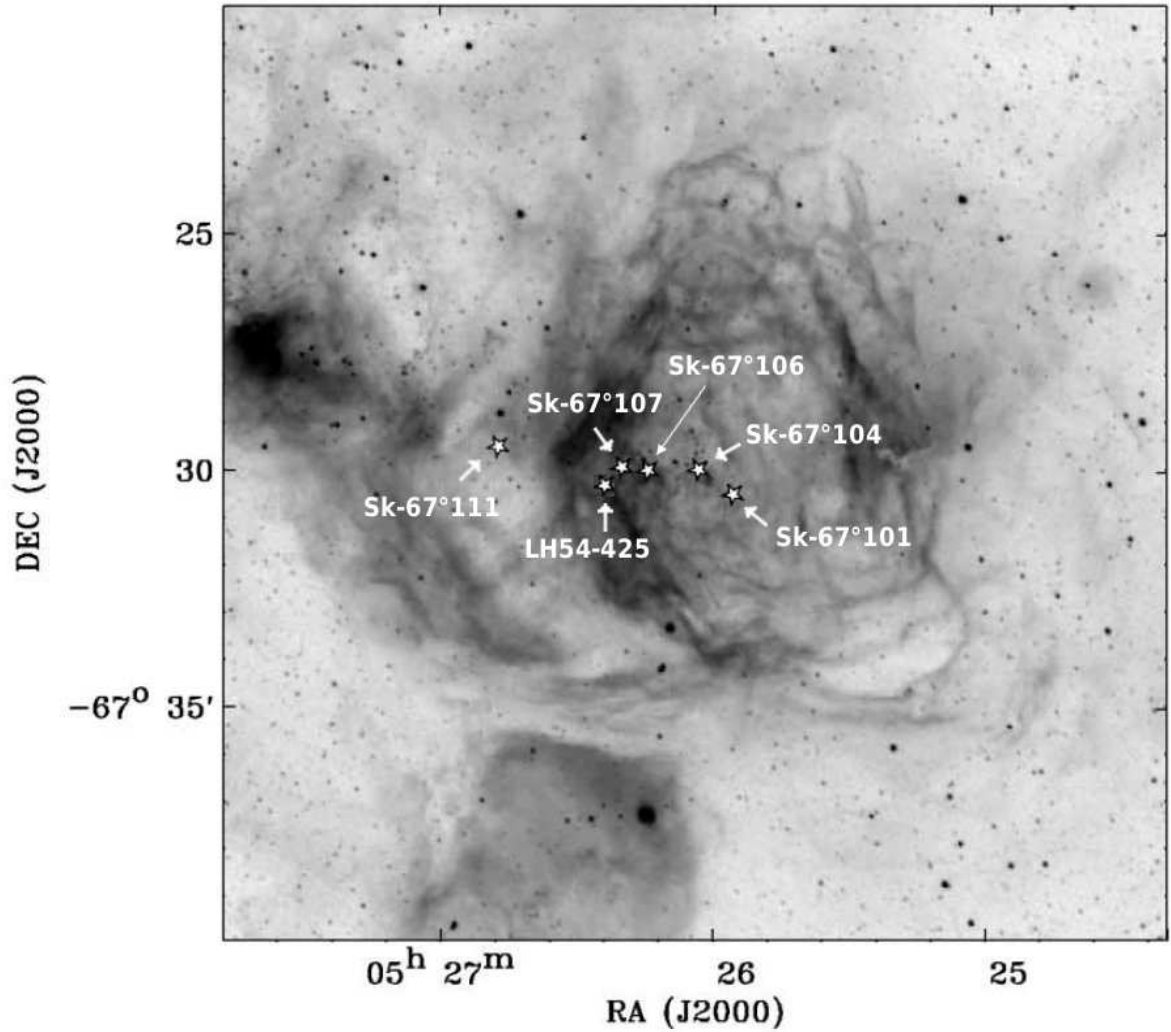} 
\caption{H$\alpha$ image of six background stars (star symbols) in their LMC environment with their Dec (in degrees) and RA (in hours) coordinates. Note that the luminous gas around the stars is warm ionised gas of the LMC, which is not relevant for our study. (This figure was adopted from \citet{Atlas}.)} 
\label{fig:3454f1} 
\end{figure*} 
\begin{table*}[h] 
\begin{center} 
\caption{Information about the FUSE data as retrieved from the MAST archive. } 
\label{tab:Stars} 
\begin{tabular}{l l l l l l l l l} 
\hline\hline 
Star & RA$^{\mathrm{a}}$ & Dec$^{\mathrm{a}}$  & Spectral$^{\mathrm{a}}$ & $V$ & date & programme & exp.time & aperture\\ 
     & (J2000) & (J2000) & type & [mag] & & & [ks] & \\ 
\hline 
\\ 
\eleven & 05 26 48.00 & -67 29 33.0 & O6 Iafpe & 12.57 & 2002-03-20 & C155 & 7.98 & LWRS \\ 
 & & & & & 2002-03-20 & C155 & 5.58 & LWRS \\ 
 & & & & & 2002-03-19 & C155 & 7.91 & LWRS \\ 
 & & & & & 2002-03-19 & C155 & 6.83 & LWRS \\ 
 & & & & & 2002-03-20 & C155 & 4.52 & LWRS \\ 
 & & & & & 2002-03-19 & C155 & 4.35 & LWRS \\ 
\new & 05 26 24.25 & -67 30 17.2 & O3IIIf+O5 & 13.08 & 2008-05-07 & F321 & 9.42 & LWRS \\ 
 & & & & & 2008-05-07 & F321 & 23.20 & LWRS \\ 
 & & & & & 2008-05-07 & F321 & 17.01 & LWRS \\ 
 & & & & & 2008-05-07 & F321 & 19.79 & LWRS \\ 
\seven & 05 26 24.00 & -67 30 00.0 & B0 & 12.50 & 2000-02-09 & A111 & 11.18 & MDRS \\ 
\six & 05 26 13.76 & -67 30 15.1 & B0 & 11.78 & 2000-02-09 & A111 & 11.27 & MDRS \\ 
\four & 05 26 04.00 & -67 29 56.5 & O8 I & 11.44 & 1999-12-17 & P103 & 5.09 & LWRS \\ 
\one & 05 25 56.36 & -67 30 28.7 & O8 II & 12.63 & 1999-12-15 & P117 & 1.26 & LWRS \\ 
 & & & & & 1999-12-20 & P117 & 6.25 & LWRS \\ 
 & & & & & 2000-09-29 & P117 & 4.38 & LWRS \\ 
\hline 
\end{tabular} 
\begin{list}{}{} 
\item[$^{\mathrm{a}}$]The coordinates and spectral types of the background LMC stars are from \citet{Atlas}.
\end{list}
\end{center} 
\end{table*} 
\clearpage
  
Despite the number of detections, the nature of these small-scale structures and their ubiquity is still a subject of study.  
It has not always been clear whether they reflect the variation in H\,{\sc i} column density, or whether they are caused by changes in the physical conditions of the gas over small scales.

However, some explanations have been suggested for the ISM model consisting of fine scale structures, such as filamentary structures \citep{Heiles}, fractal geometries driven by turbulence \citep{Elmegreen1997}, or a separate population of cold self-graviting clouds \citep{Walker:Wardle,Draine,Wardle:Walker}. 

In absorption line spectroscopy most cases of temporal or spatial small-scale structures are observed in optical high-resolution spectra which, however, often provide little information on the physical conditions in the gas. 
Using UV spectra, valuable information can be gained about the physical properties of the gas. From fine structure levels of \ion{C}{i}, \citet{Welty2007} discovered that the observed variations in Na\,{\sc i} and Ca\,{\sc ii} lines towards HD\,219188 reflect the variation in the density and ionisation in the gas, and do not refer to high-density clumps.
A different approach was used by \citet{Richter:2003b,Richter:2003a} who found, using the many absorption lines of molecular hydrogen in the Lyman and Werner bands, that the molecular gas in their studied lines of sight resides in compact filaments, which possibly correspond to the tiny-scale atomic structures in the diffuse ISM \citep{Richter:2003b}. 
 
The Large Magellanic Cloud (LMC) has many bright stars with small angular separation on the sky, and provides a good background for studying the gas in the 
disc and the halo of the Milky Way at small scales. Here we have chosen the six 
LMC stars \eleven, \new, \seven, \six, \four, and \one, with small angular 
separations to study the structure of that foreground gas. We refer the interested reader to \citet{Atlas} for an atlas of FUSE spectra of the LMC sight-lines. 
 
This paper is organised as follows: 
In Sect.\,\ref{sect:Data} we describe the data used in this work.  
In Sect.\,\ref{sect:Handling of spectral data} we explain the LMC sight-line and the methods used to gain information from the spectra. 
In Sect.\,\ref{sect:H2} and Sect.\,\ref{sect:Metal} respectively we discuss the results for molecular hydrogen and metal absorption. 
In Sect.\,\ref{sect:Density} we derive a density relation based on H$_2$ and O\,{\sc i} to investigate the density variations within the sight-lines, and combine that with the derived average density from the C\,{\sc i} excitation to understand the properties of the gas. 
We discuss the results and interpretations in Sect.\,\ref{sect:summary}.

\section{Data}

\label{sect:Data} 

We used far ultraviolet (FUV) absorption line data from the Far Ultraviolet Spectroscopic Explorer (FUSE) with relatively high S/N to analyse the spectral 
structure in six lines of sight with small angular separations. For four of our sight-lines high-resolution Space Telescope Imaging Spectrograph (STIS) observations were also available. In total we cover the wavelength range 905-1730\,{\AA}, where we can find most electronic transitions of many atomic species and molecular hydrogen. 

\begin{table}[tbh] 
\begin{center} 
\caption{S/N per resolution element measured within a few {\AA}ngstr{\"o}m around the listed wavelength.} 
\label{tab:SN} 
\begin{tabular}{l c c c c c c} 
\hline\hline 
Star & LiF:1A & 2A  & 1B & 2B & SiC & STIS \\ 
 & \hspace{-10mm}$\lambda$[{\AA}]= 1047 & 1148 & 1148 & 1047 & 980 & 1605 \\ 
\hline 
\eleven & 97 & 74 & 66 & 71 & 55 & - \\ 
\new & 88 & 68 & 60 & 58 & 44 & - \\ 
\seven & 42 & 41 & 35 & 26 & 17 & 17 \\ 
\six & 62 & 55 & 48 & 30 & 16 & 17 \\ 
\four & 57 & 50 & 42 & 35 & 31 & 21 \\ 
\one & 44 & 52 & 40 & 36 & 20 &  16 \\ 
\hline 
\end{tabular} 
\end{center} 
\end{table} 

The FUSE instrument consists of four co-aligned telescopes and two micro-channel plate detectors, one coated with Al+LiF, the other with SiC. Each has their maximum efficiency in different parts of the spectral range. The spectral data have a resolution of about 20 \kms\,(FWHM), and cover the wavelength range 905-1187\,{\AA}, of which 905-1100\,{\AA} are by the SiC coatings and 1000-1187\,{\AA} by the Al+LiF coatings. Each detector is divided into two segments, A and B. 
For detailed information about the instrument and observations see \citet{Moos} and \citet{Sahnow}.  
 
In the order of right ascension our target stars are: \eleven, \new, \seven, \six, \four, and \one\ (see Fig.\,\ref{fig:3454f1}). These are 
all bright stars of spectral types O and B, lying on an almost straight line at 
constant declination, and spread over $5\arcmin$ in RA. Because the stars are separated mainly in 
their RA, we have considered the projected separations in that coordinate. 
The smallest separation is however between \new\ and \seven, which has almost 
the same RA, which means that we have to take into account the true separation of $17\farcs2$ (essentially in the declination), when comparing these two sight-lines. 

We retrieved the already reduced data from the MAST\footnote[1]{Multi-Mission Archive at Space Telescope: http://archive.stsci.edu/} archive and co-added them separately for each of the four co-aligned telescopes. For the information about the retrieved data as well as the properties of the stars, see Table\,\ref{tab:Stars}. The data were all reduced with the {\sc calfuse} v3.2.1, except for the \new\ and \six, which were reduced with {\sc calfuse} v3.2.2. The typical S/N of the 
data varies for different detector channels and sight-lines (see Table\,\ref{tab:SN} for more information on the S/N of the FUSE and the STIS data). The data from the SiC channels have in general lower S/N, while the data quality is significantly higher for all the FUSE detectors for the \eleven\ 
and \new\ sight-lines.  
Due to the different wavelength zero points in the calibrated data, the spectra of different stars and different cameras might shift within $\pm20$\,\kms. This shift however is steady for the whole spectral range and does not affect the line identification. 
 
The STIS observations were part of a programme for studying the N\,51\,D superbubble (Wakker et al., in preparation). Only the stars \seven, \six, \four, and \one\ were observed in that programme. STIS provides a better resolution of $6.8$\,\kms, and covers the wavelength range longward of the FUSE spectrum from about 1150 to 1730\,{\AA}. 

\begin{figure*}[t] 
\centering 
\includegraphics[scale=0.93]{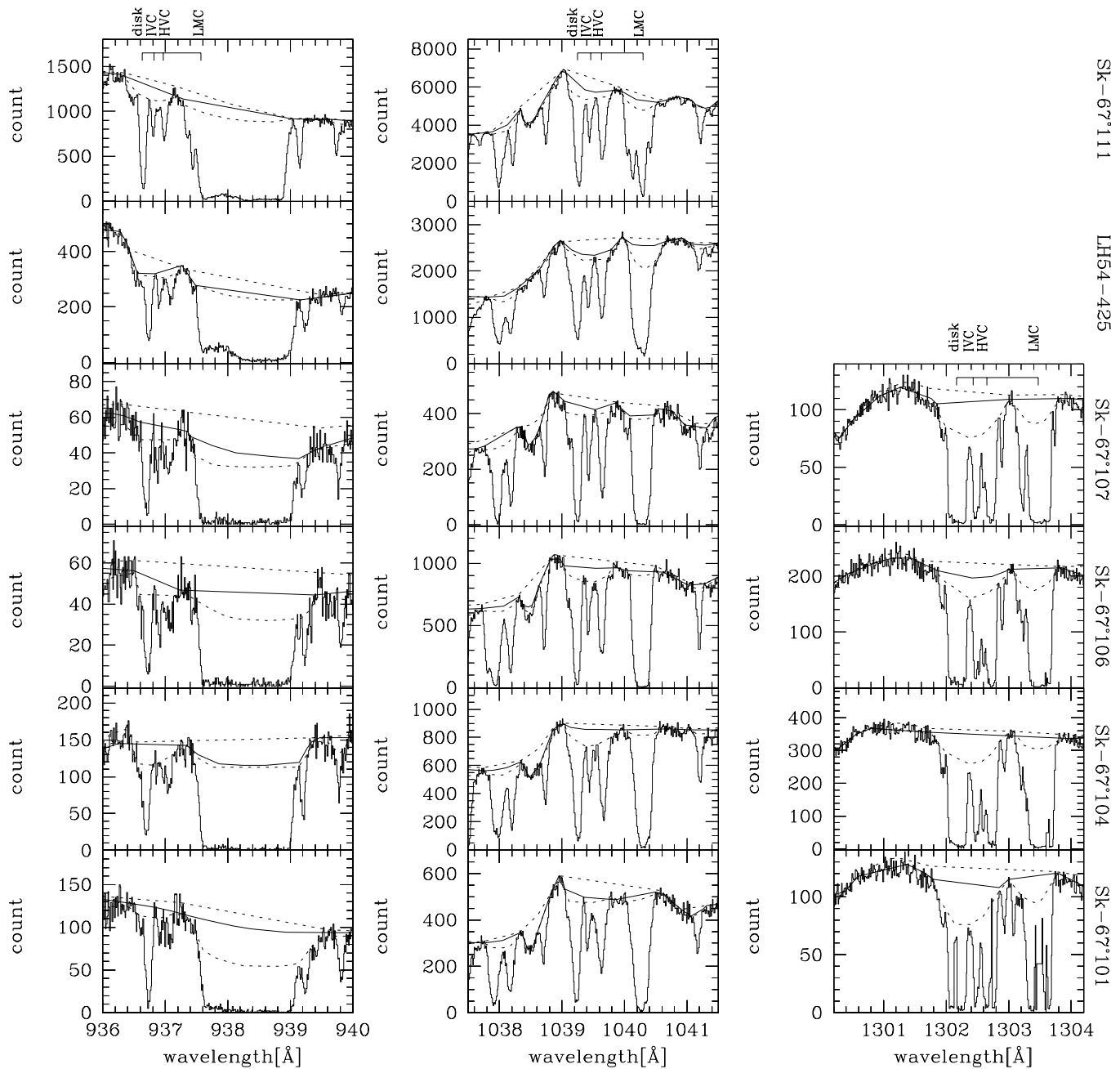} 
\caption{O\,{\sc i} absorption at 936.63 (from FUSE:SiC), 1039.23 (from FUSE:LiF), and 1302.17\,{\AA} (from STIS) and the continuum fit is shown for all six stars (names given at right). The solid line is the continuum fit to which the spectrum has been normalised, the dashed lines show the upper and lower continuum choice. The bars at the top mark the known absorption component velocities for the O\,{\sc i} lines (see Sect.\,\ref{subsect:LMCline}).} 
\label{fig:3454f2} 
\end{figure*} 

\section{Handling of spectral data}

\label{sect:Handling of spectral data}

\subsection{LMC sight-line} 
\label{subsect:LMCline} 

The line of sight towards the LMC is rather complicated, because it contains several gas 
components. Knowing the absorption lines in the ISM with their transition 
wavelengths we are able to identify the different elements, and distinguish the 
different gas components through their radial velocities. 
 
We followed the comprehensive description given by, e.g., 
\citet{Savage:deBoer}. On a line of sight there is gas in the solar vicinity, 
normally seen around $v_{\rm rad} \simeq 0$ \kms\ (all velocities given are LSR 
velocities). Next is the intermediate velocity gas around $\simeq +60$ \kms\ on 
essentially all LMC lines of sight; this gas likely is a foreground intermediate velocity cloud (IVC). Most lines of sight also show the presence of a 
high-velocity cloud (HVC) at $v_{\rm rad} \simeq +110$ \kms. The properties of the IVC and HVC will be discussed in Nasoudi-Shoar et al. (in preparation). Finally, the LMC 
gas itself shows its presence in the velocity range between +180 and +300 \kms.  
In this paper we analyse the Milky Way disc component at the velocity $v_{\rm rad}=16\pm 0.5$ \kms, where the various rotational levels of molecular hydrogen have relatively strong absorption and allow for a detailed study of the properties of the gas. The absolute wavelength calibration of the STIS data is much more reliable, therefore we base our radial velocities on these data. 
This reveals an additional weak component at $v_{\rm rad}=-23\pm 1.8$ \kms, which we describe briefly, while we concentrate on the analysis of the strong disc component. 
We refer to the absorptions with similar radial velocity as parts of the same cloud.

\subsection{Column-density measurements} 
 
The continua of the spectra consist of the spectral structures from the 
background stars with broad lines and P-Cygni profiles.  
Before measuring the equivalent widths we estimated a continuum level by eye, and normalised the spectra to 
unity with the programme {\sc spectralyzor} \citep{Ole}. In this way we eliminated the large scale variation of the continuum as well as the possible variations on scales less than 1{\AA}. 
The absorption lines from those parts of the spectra where the continuum placement is highly uncertain were considered with special caution or were excluded from the final analysis. 
 
For each velocity component the equivalent width of the absorption was measured 
either by integrating the pixel to pixel area up to the continuum level, or by Gaussian fits with {\sc midas alice} (only for unsaturated lines).
Owing to the large number of spectral lines within the FUSE wavelength range and the wide velocity absorption range for each transition, there is considerable blending in the interstellar lines. For the absorption lines that needed deblending, a multi-Gaussian fit was used, except for the clearly saturated lines, where the former method was more suitable. 
For the final results we only took those lines into consideration where we found no blending or which could easily be decomposed by a multi-Gaussian fit. For the \ion{O}{i} line at 1302{\AA} however, which is heavily saturated and also blended with its also saturated component at $-23$\,\kms, we used a different approach. We used profile fitting in addition to the curve-of-growth (\cog) technique, with the derived column density and $b$-value from the other \ion{O}{i} lines (based on the \cog\ method) given as the initial guess, to separate the equivalent widths of the two components. 
 
The dominant source of error in equivalent-width measurements is the placement of the continuum.  
We therefore estimated the errors based on the photon noise and the global shape of the continuum.  
The errors due to the continuum fit were measured by manually adopting a highest/lowest continuum within a stretch of few {\AA}ngstr{\"o}m around each absorption line. 
The minimum errors in this way depend on the local S/N and correspond to the change of the continuum around $\pm 1\sigma$ noise level.  
Due to the complex shape of the stellar continuum, this way of choosing the continuum by eye is more reliable than an automatic continuum fitting. Examples are shown in Fig.\,\ref{fig:3454f2} for some of the \ion{O}{i} lines. 
  
We selected primarily absorption lines of H$_2$ to study the fine structure of the gas. This allows us to derive the physical properties of the gas. 
Furthermore, H$_2$ exists in the coolest portions of the 
interstellar clouds and thus will produce the narrowest absorption lines. 
Moreover, in the FUV range many absorption lines from the same lower 
electronic level are available, thus allowing the easy determination of the 
respective \cog\ (see, e.g., \cite{deBoer98}, \cite{Richter98}, \cite{Richter00}). 
We also included available metal lines in the study, which are the lines of \ion{C}{i}, \ion{N}{i}, O\,{\sc i} , Al\,{\sc ii}, Si\,{\sc ii}, P\,{\sc ii}, S\,{\sc iii}, Ar\,{\sc i}, and \ion{Fe}{ii}. 
For the four sight-lines with available STIS observations, the lines of \ion{C}{i*}, \ion{C}{i**}, Mg\,{\sc ii}, Si\,{\sc iv}, S\,{\sc ii}, Mn\,{\sc ii}, and Ni\,{\sc ii} were additionally detected. 
 
A standard \cog\ method was used to estimate the column density $N$ and the 
Doppler parameter $b$ for each species. 
The theoretical {\sc CoG}s were constructed for a range of $b$-values in the 
interval of 1\,\kms, based on the damping constant of the strongest measured transition in each species. The column densities were determined by finding the the best representative \cog\ for the set of measured equivalent widths and their known log($f\lambda$).
The $f$-values used are taken from the list of \cite{Abgrall:a,Abgrall:b} for H$_2$, and \citet{Morton} for other species. The final column densities were derived through a polynomial regression and are presented in Table~\ref{tab:N} together with the $1\sigma$ errors\footnote[2]{Based on 68$\%$ of the data points having a residual within $1\sigma$ from the corresponding \cog\ coefficient.}. 
Thus the uncertainties in the column densities are based on the statistical errors in equivalent-width measurements and the uncertainties in the $b$-values.  

\section{Molecular hydrogen in Galactic disc gas} 
 
\label{sect:H2} 

\begin{figure}[] 
\centering 
\includegraphics[scale=0.45]{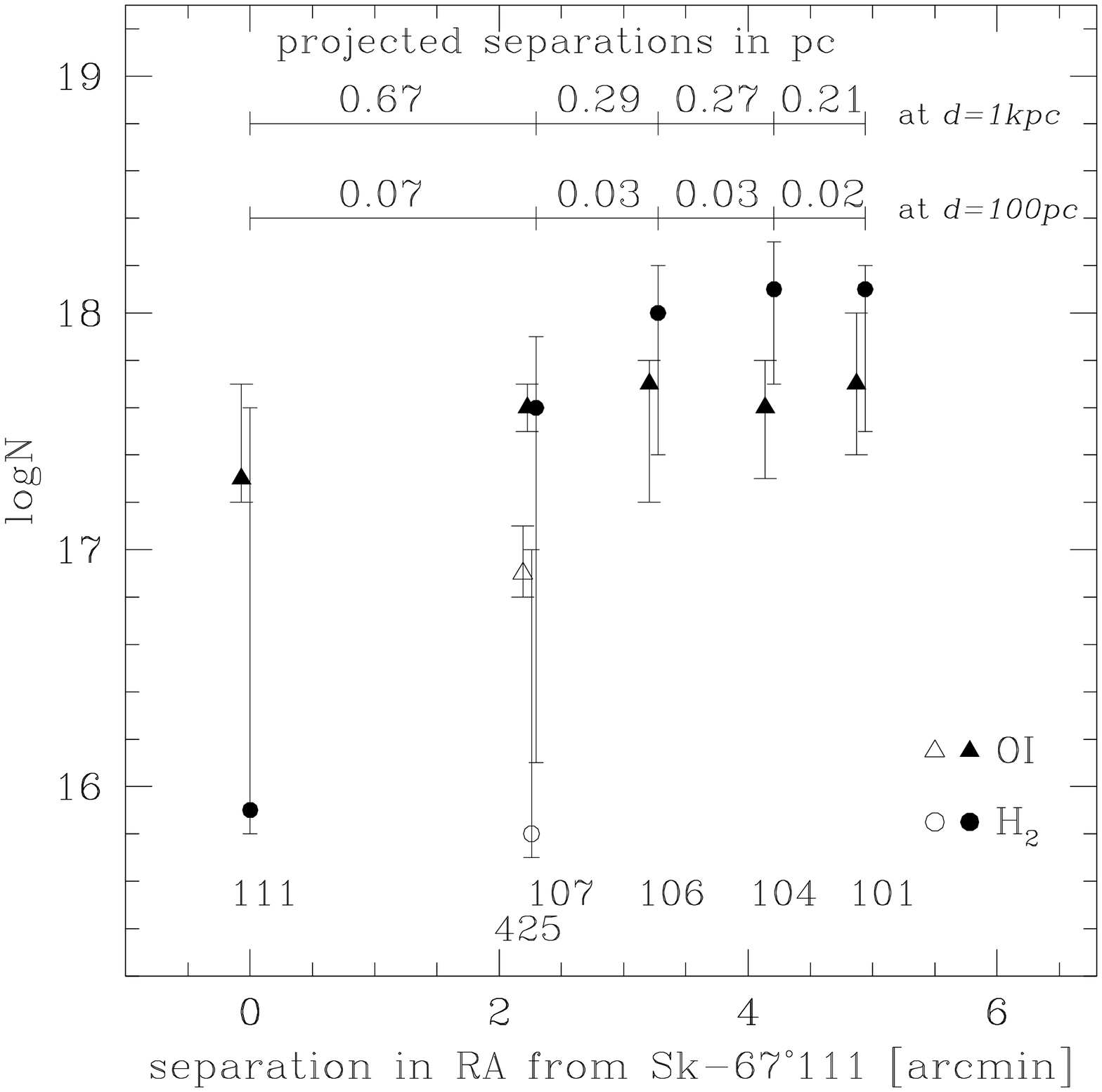} 
\caption{Logarithmic column densities of H$_2$ and O\,{\sc i} towards \eleven, \new, \seven, \six, \four\, and \one\ (from left to right; for the identification see the numbers in the lower part of the plot), plotted against the angular separation of each star with respect to \eleven\ (only in RA). 
The data for all sight-lines are presented as filled symbols (circles for H$_2$, and triangles for \ion{O}{i}), except for \new, where they are given as empty symbols.
The O\,{\sc i} data were slightly shifted to the left for a better visualisation. Data for \new\ were shifted slightly more to the left for clarity. 
On top of the plot the projected linear separations in pc between each pair of 
stars is marked for the assumed distances of $d=100$\,pc respective $d=1$\,kpc. The increase of the H$_2$ column density from \eleven\ to \one\ is not seen in the O\,{\sc i} column density.} 
\label{fig:3454f3} 
\end{figure} 

We determined the column densities of the molecular hydrogen for the different rotational states $J=0-4$. These are given in logarithmic values in Table\,\ref{tab:N} together with the corresponding $b$-values.  
The $b$-values for various rotational states of molecular hydrogen are estimated 
separately in each \cog\ analysis and vary over the range of 2-6\,\kms\ for the low $J$ for all 
sight-lines. Thus, the various $J$-levels might differ slightly in their 
$b$-values, because they may sample different physical regions of the Galactic disc gas.  
 
In our FUSE spectra we did not detect any level higher than $J=4$ for any of our sight-lines. Moreover the $J=4$ detection was not always certain. We therefore estimated a 1\,$\sigma$ equivalent width $EW_{1\sigma}$ based on the local noise fluctuation and the FUSE spectral resolution: $EW_{1\sigma}=1.06\times FWHM_{\rm inst}\times \sigma_{\rm noise}$. For sight-lines with no measured absorption above $3\sigma$, we used a $EW_{3\sigma}$ based on the strongest unblended transition in a fit to the linear part of the \cog\ for estimating an upper limit for the column densities\footnote[3]{The fit has been including all possible $b>2$ \kms.}. 
 
In order to visualise the various regions of the foreground cloud we are looking at, these column 
densities are plotted in Fig.\,\ref{fig:3454f3} against the angular separation of the 
background stars. Assuming that the gas exists in the disc at distances up to 
1\,kpc (Lockman et al. 1986), we have marked in that figure the projected separations between our 
sight-lines. The H$_2$ absorption shows a clear decline in column density 
towards \eleven, with a significant drop towards \new.

\subsection{Note on the errors} 
 
\begin{figure*}[tbp] 
\centering 
\includegraphics[scale=0.93]{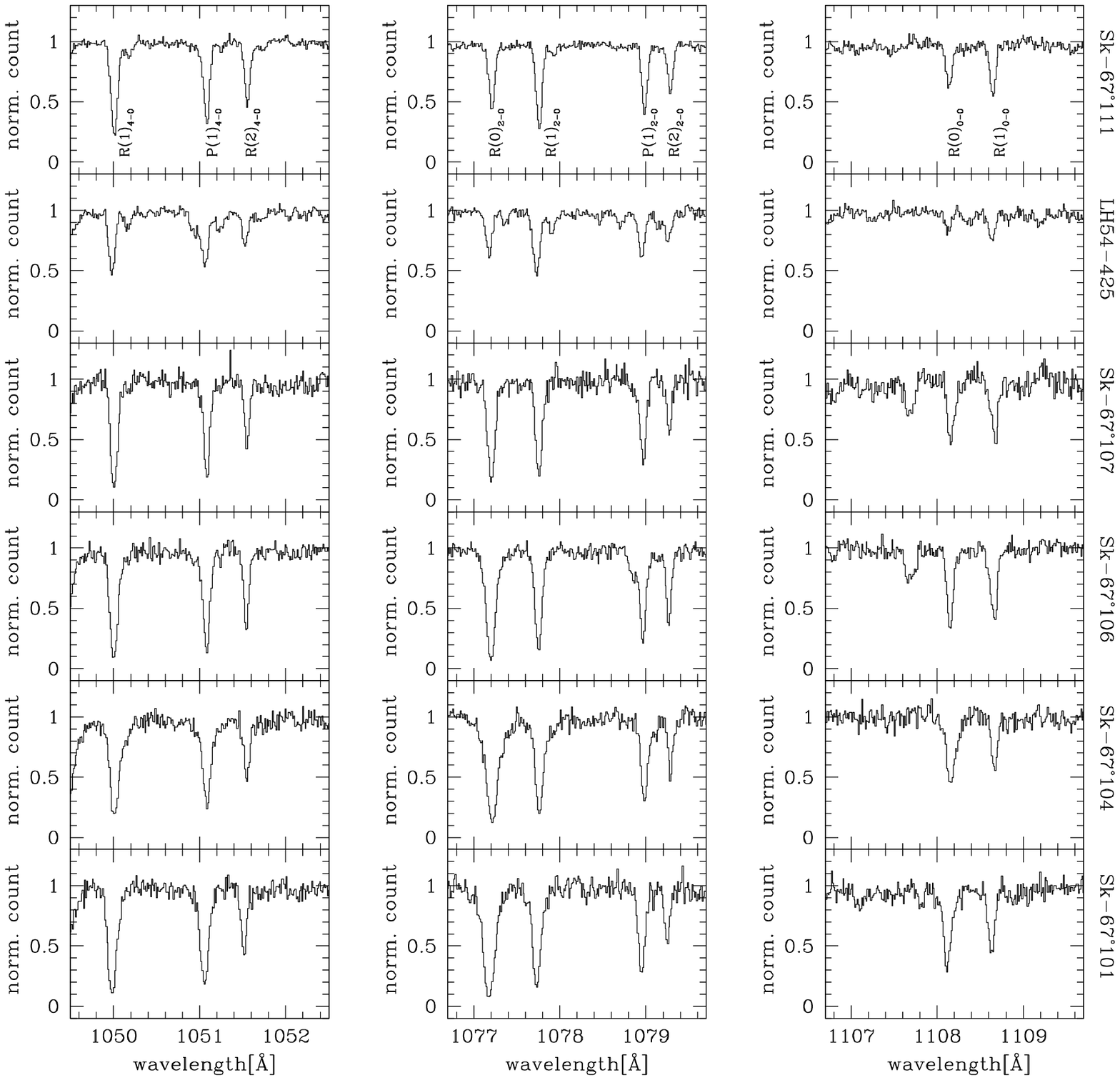} 
\caption{Sample of H$_2$ absorption lines from different excited levels shown for the different sight-lines. The sight-lines are labelled on the right, from top to bottom: \eleven, \new, \seven, \six, \four, and \one. The spectra are normalised to unity in all cases. The corresponding transitions are labelled on the upper plots of \eleven, the same portion of the spectrum is plotted in each panel for the other sight-lines. 1\,{\AA} corresponds to 270 \kms.} 
\label{fig:3454f4} 
\end{figure*}

There are different sources of errors contributing to our measured column densities. 
We included the statistical errors due mainly to the continuum placement and the local S/N in the equivalent-width measurements. However, the different detectors seem to have different responses, sometimes causing slightly lower or higher equivalent widths of the same absorption. This in itself should be included in our errors in the column densities, but combining the measurements from different detectors might influence the fit to the \cog, because they include different transitions in a wide range on the \cog.  
For the $J=1$ lines, a crucial role is played by one line at 1108\,{\AA}. If that line is for some reason an outlier on the \cog, $N(J=1)$ could be overestimated for the \one\ to \seven\ sight-lines. 
On the other hand, a direct comparison of the absorption profiles clearly shows that the components towards \new\ are much weaker than on the other sight-lines, particularly for $J=0$ and $J=1$ (see Fig.\,\ref{fig:3454f4} for some examples). 
Furthermore, it is also obvious that the $J=0$ lines towards \eleven\ are much weaker than the $J=1$ lines in comparison to the other sight-lines. Thus, the differences in the $N(J=0)$/$N(J=1)$ ratios between the different sight-lines are real.  

The large errors in the derived column densities for some of the $N(J=0)$ and $N(J=1)$ are mostly due to a degeneracy of the $N$-$b$ combinations, rather than covering the whole range of possible column densities; i.e. given that our best fit represents the ``true'' $N$-$b$ combination, the errors should be much smaller than the indicated errors. 

\subsection{Physical properties of the gas from H$_2$} 

The various rotational levels of molecular hydrogen reveal the physical state of the gas. 
While lower $J$-levels are mostly populated by collisional excitations, higher 
$J$-levels are excited by photons through UV-pumping 
\citep{Spitzer+Zweibel}. The population of the lowest levels can be fitted to a 
Boltzmann distribution resulting in $T_{\rm exc}$ for the gas, the population of 
higher levels can be fitted in a similar way leading to an $"$equivalent 
UV-pumping temperature$"$, $T_{\rm UV-pump}$.  

\begin{figure*}[tbp] 
\centering 
\includegraphics[scale=0.9]{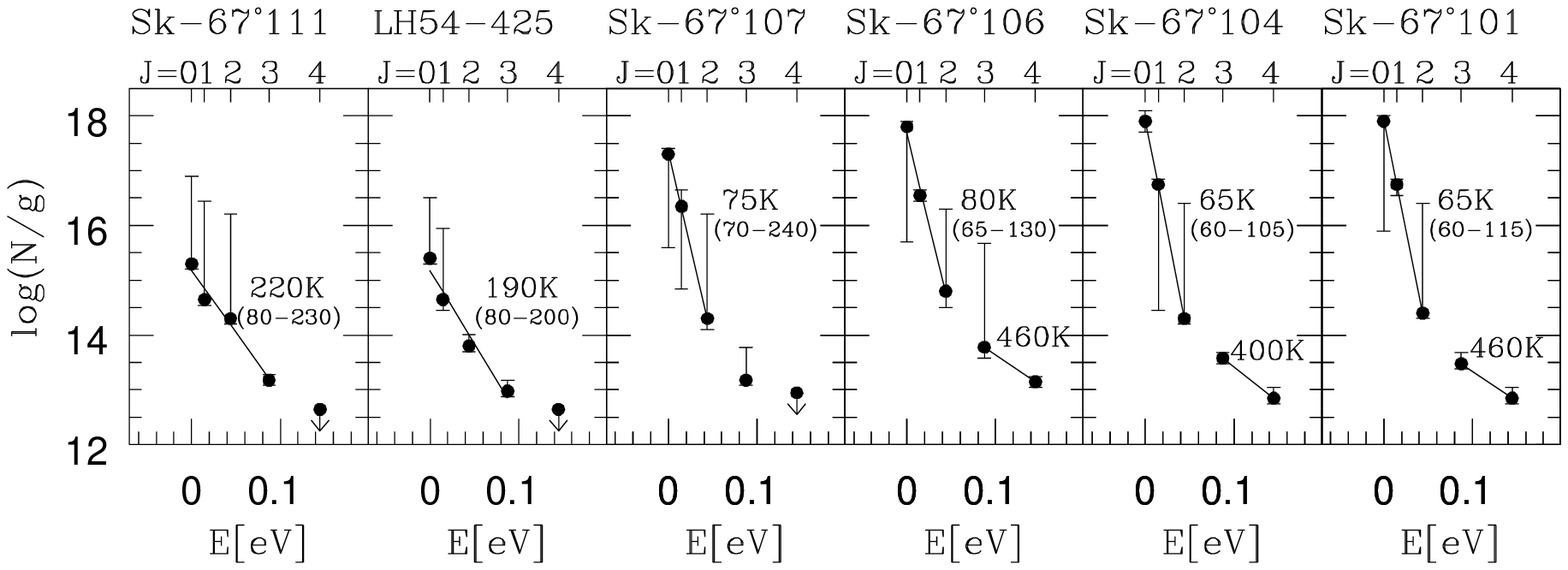} 
\caption{Rotational excitation of H$_2$ in foreground Galactic disc gas. For each sight-line the column density [cm$^{-2}$] of H$_2$ in level $J$ is 
divided by the statistical weight $g_J$ and plotted on a logarithmic scale 
against the excitation energy $E_J$. The corresponding Boltzmann temperatures are given with the fitted lines. 
This temperature is represented by a straight line through the three lowest rotational states $J=0$, 1, and 2. The higher $J$-levels indicate the level of UV-pumping. For \eleven\ and \new, $T_{0,3}$ is determined instead, because the higher $J$-levels as well as the lower ones may be populated by UV-pumping. The values within parenthesis are the upper/lower limits for the temperature based on the extremes of the errorbars (see Sect.\,\ref{sect:H2}.1).} 
\label{fig:3454f5} 
\end{figure*}

In Fig.\,\ref{fig:3454f5} we plotted the column density of H$_2$ in level $J$, divided by the statistical weight, $g_J$, against the rotational excitation energy, $E_J$. 
For most of our sight-lines the rotational excitation can be represented by a two-component fit. A Boltzmann distribution can be fitted to the three 
lower levels $J=0$, $J=1$, and $J=2$ to find the excitation temperature 
$T_{0,2}$ of the gas. We get the equivalent UV-pumping temperature $T_{3,4}$ through a fit 
to the $J=3$ and $J=4$ levels.  
The Boltzmann excitation temperature ranges from 65\,K towards \one\ and \four, 
to about 80\,K in the directions of \six\ and \seven. This is in the typical temperatures range found for the cold neutral medium, CNM. 
  
Towards \new\ and \eleven, on the other hand, a fit to the $J=0-3$ levels would give a $T_{0,3}$ similar to $T_{0,1}$ and $T_{0,2}$, indicating that the $J=2$ level is already excited by UV-pumping. 
Given that the derived column densities for $J=4$ are upper limits, the gas in these two lines of sight may be fully thermalised, and a single Boltzmann fit through all points might be sufficient.  
In this case we have derived the excitation temperature $T_{03}\sim$200\,K. 

Furthermore, the temperature plot confirms the estimated column densities, as the indicated errors by the \cog\,-fit would lead to unrealistic ratios of the column densities and the statistical weight between the different $J$-levels. 
This steep drop of $N(J)$ from $J=0$ to $J=2$ is also generally seen in the ISM for sight-lines with $\log N(J=0)>17$ \citep{Spitzer:Cochran}. 

Molecular hydrogen in the interstellar medium forms in cores of clouds where the gas is self-shielded from the ultraviolet radiation. The two-component fit to four of our sight-lines with low $T_{\rm exc}$ can be interpreted as a core-envelope structure of the partly molecular clouds, where the inner parts are self-shielded from the UV radiation.  
Detecting less H$_2$ towards \eleven\ and \new, together with the higher $T_{\rm exc}$ in these two directions, can be understood as looking towards the edge of the H$_2$ patch in those sight-lines, where the H$_2$ self-shielding is less efficient.  
However, \seven, with a true separation of only $17.2\arcsec$ from \new, 
shows a significantly higher column density of H$_2$ and a much lower temperature. Assuming that the disc gas exists at a distance between 100\,pc to 1\,kpc, the projected linear separation between \new\ and 
\seven\ corresponds to a spatial scale of 0.01 to 0.08\,pc (essentially only in declination). This suggests that the observed H$_2$ patch has a dense core and a steep transition to the edge. 

\section{Metal absorption and abundances} 
 
\label{sect:Metal} 

We determined the column densities for the species \ion{C}{i}, \ion{N}{i}, O\,{\sc i}, Al\,{\sc ii}, Si\,{\sc ii}, P\,{\sc ii}, S\,{\sc iii}, Ar\,{\sc i}, and \ion{Fe}{ii} from the FUSE spectra. In addition, the column densities of \ion{C}{i*}, \ion{C}{i**}, Mg\,{\sc ii}, Si\,{\sc iv}, S\,{\sc ii}, Mn\,{\sc ii}, and Ni\,{\sc ii} were obtained for \one, \four, \six, and \seven\ using STIS data.  
These metal column densities are presented in Table\,\ref{tab:N}. For a list of absorption lines used for the column density determination, see Table\,\ref{tab:lines}. Some of the species in the FUSE spectral range have additional transitions in STIS spectra, which made a more accurate determination of their column densities possible. This, together with the location of the data points on the \cog, results in uncertainties in the column densities that are smaller along some sight-lines compared to others. 

The better resolution of the STIS data allows us to separate the velocity component at $\sim -20$\,\kms\ that otherwise appears as an unresolved substructure in, e.g., \ion{P}{ii} and O\,{\sc i} absorption in the FUSE spectra. This component is strongest in the metal lines of O\,{\sc i}, Mg\,{\sc ii}, and Si\,{\sc ii}.
We do not consider this component any further, as it may be part of an infalling cloud, which is not connected to the disc gas for which we study the small-scale structure in this paper. Only for strong saturated lines does this component affect our measurements, as we mentioned in Sect.\,\ref{sect:Handling of spectral data}. 

Within our spectral range \ion{Fe}{ii} appears in a number of transitions with 
a wide range in $f$-values, making an accurate fit to the \cog\ possible. 
Some of the other species, on the other hand, have only few transitions in our spectra within a small ${\rm log}f\lambda$ range, and therefore can be fit to a wide range of $b$-values.   
Owing to the lack of more information about the Doppler widths of these species, we adopted the $b$-value of \ion{Fe}{ii} for each line of sight, and 
we determined the column densities of other ions from the \ion{Fe}{ii}-\cog. There are some exceptions discussed below for \ion{O}{i}, \ion{N}{i}, and \ion{Ar}{i} (and \ion{C}{i}, discussed in the next section).  

\begin{figure*}[th]
\centering 
\includegraphics[scale=0.88]{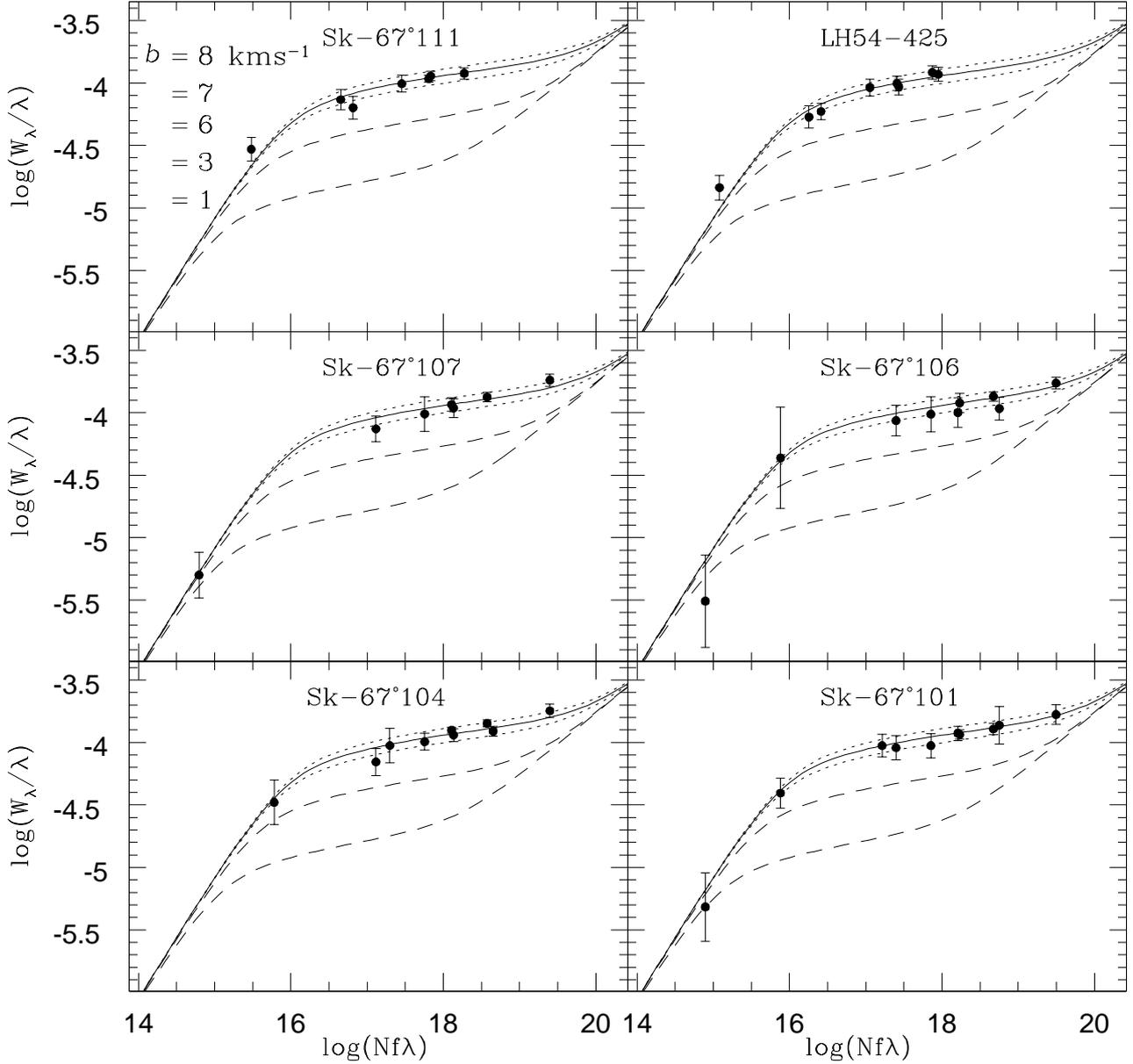} 
\caption{Sample of {\sc CoG}s of \ion{O}{i} for the six sight-lines (the star name on top of each figure). Each plot shows {\sc CoG}s for $b=1,3,6,7,8$\,\kms. The solid line is the resulting \cog\ from the best fit ($b=7$\,\kms), and the dotted lines ($b=6$ and $b=8$\,\kms) correspond to approximately $1\sigma$ errors.} 
\label{fig:3454f6} 
\end{figure*}

For \ion{O}{i} we find a number of transitions within the FUSE spectral range, and two additional lines within the STIS spectral range (see Table\,\ref{tab:lines}). 
The derived Doppler parameters of \ion{O}{i} are very similar to those of \ion{Fe}{ii}. This can be expected because turbulent broadening dominates thermal broadening in the LISM. 
We have derived them separately however. 
With both saturated and very weak \ion{O}{i} lines, we are able to make a relatively accurate fit to the \cog\ (see Fig.\,\ref{fig:3454f6}) and do not need to rely on the accuracy of the \ion{Fe}{ii}-\cog. Adopting the \ion{Fe}{ii}-\cog\ would furthermore lead to unsatisfying fits in most cases. 
We plotted in Fig.\,\ref{fig:3454f3} the measured O\,{\sc i} column densities together with those of H$_2$. The $N\rm (\ion{O}{i})$ stays constant within the errors between the lines of sight, except for the decrease towards \new. 
 
\ion{N}{i} is also detected in a number of transitions. These lines also yield $b=7$\,\kms, even though they are mainly on the flat part of the \cog, which makes the determination of the \cog\ less certain.  
\ion{N}{i} probably originates from the same physical region of the neutral gas as \ion{O}{i}. We therefore adopted the \ion{O}{i}-\cog\ to determine the column densities of \ion{N}{i}. 
Assuming that also \ion{Ar}{i} mainly exists with the \ion{O}{i} in the neutral gas, we fitted the two detected \ion{Ar}{i} absorptions to the \ion{O}{i}-\cog\ as well.

\ion{Si}{ii} has only one moderately strong line in the FUSE spectra (see Table\,\ref{tab:lines}) and some very strong absorptions in STIS. The absorption at 1020.7 {\AA} is kind of an outlier though when adopting the \ion{Fe}{ii}-\cog, because it appears too low on the \cog. Hence the reduction of the $N(\ion{Si}{ii})$ towards \eleven\ and \new, compared to the rest of the sight-lines, is mainly due to the lack of further available data, rather than reflecting the differences in the equivalent width of the 1020.7 line. 
We therefore represent the $N(\ion{Si}{ii})$ towards  \eleven\ and \new\ as lower limits in Table\,\ref{tab:N}. It is possible that this problem occurs because of the sampling of different gas components. This would appear more strongly in the stronger absorptions compared to the weaker ones, which would explain why the 1020.7 line appears too low on the \cog, compared to the much stronger lines of the STIS spectra.

\begin{figure}[tbp]
\centering 
\includegraphics[scale=0.43]{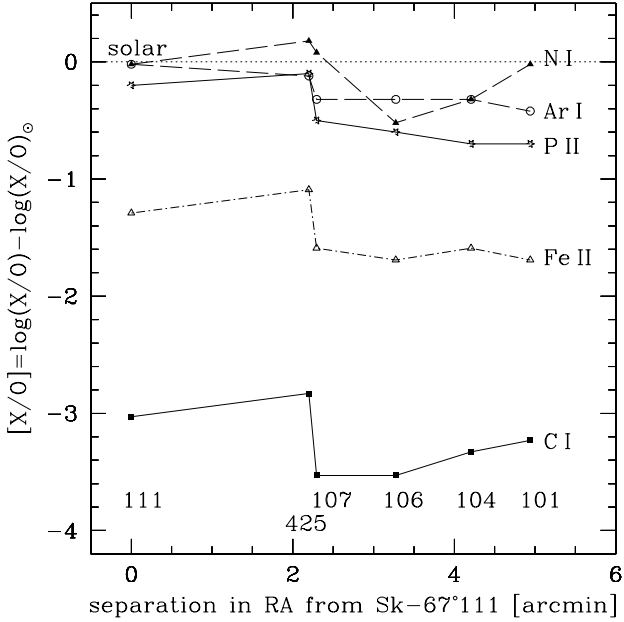} 
\caption{Metal abundances plotted against the angular separation. The abundances are expressed as the ratio $\rm 
[X/O]=log(X^{i}/O)-log(X/O)_{\odot}$, assuming that the $X^{i}$ is the dominant 
ion of $X$. $X^{i}$ is given on the right side of the figure.} 
\label{fig:3454f7} 
\end{figure}

The metal column densities show mostly insignificant variations between 
our lines of sights, but are generally the lowest to \new. Below we investigate the variations in form of abundance ratios. 
Due to the lack of information about H\,{\sc i} column densities with a spatial 
resolution that corresponds to our particular sight-lines, we derive the 
abundances as the ratio $\rm [X/O]=log(X^{i}/O)-log(X/O)_{\odot}$, with the 
solar abundances taken from \citet{Asplund}, and the assumption that the 
detected $X^{i}$ is the dominant ion of the element $X$. O\,{\sc i} is an ideal 
tracer for H\,{\sc i} since, unlike iron and silicon, only small fractions are depleted 
into dust grains, it has the same ionisation potential as hydrogen, and both 
atoms are coupled by a strong charge-exchange reaction in the neutral ISM. 
Some of the derived metal abundances are plotted in Fig.\,\ref{fig:3454f7} against the angular separation of the sight-lines. 

\ion{N}{i}, with an ionisation potential of 14.1\,eV, acts as another tracer for the neutral gas. Our derived [\ion{N}{i}/\ion{O}{i}] abundances are in general close to solar, except towards \six\ and \four, where they are somewhat lower. 
The ionisation balance of \ion{N}{i} is more sensitive to local conditions of the ISM, as \ion{N}{i} is not as strongly coupled to \ion{H}{i} as \ion{O}{i} \citep{Jenkins2000,Moos2002}. We therefore can expect the [\ion{N}{i}/\ion{H}{i}] to vary within a spatial scale, while [\ion{O}{i}/\ion{H}{i}] remains constant.  
Thus, the variation of [\ion{N}{i}/\ion{O}{i}] could suggest a lower shielding towards the sight-lines with observed lower [\ion{N}{i}/\ion{O}{i}]. \cite{Jenkins2000} have modelled the deficiency of local \ion{N}{i} with decreased \ion{H}{i} column density. It is thus possible that the variation of [\ion{N}{i}/\ion{O}{i}] between our sight-lines are caused by the actual variations in the density.  
The \ion{N}{i} absorptions are, however, not easily measured due to the multiplicity of the transition, which can lead to high uncertainties in the \ion{N}{i} column densities. 
 
Fe, Mn, and Ni are generally depleted into dust grains in the Galactic disc, which can explain their low relative abundances (Fig.\,\ref{fig:3454f7}). 
The underabundance of Ar can, however, be explained by the large photo-ionisation cross section of \ion{Ar}{i} \citep{Sofia}. 
The slightly higher [Ar/O] towards \new\ and \eleven\ would then suggest a higher shielding in these sight-lines, compared to the others, given that the derived column densities are precise.  
[\ion{Si}{ii}/\ion{O}{i}], on the other hand, is close to solar in the lines of sight \seven\ to \one\ (with available weak and strong absorption lines).
As \ion{Si}{ii} is the dominant ion in both warm neutral medium (WNM), and warm ionised medium (WIM), these ratios might indicate that part of the \ion{Si}{ii} absorption originates from the ionised gas. If silicon is partly depleted into dust grains, these ratios could be even higher. 
Furthermore, the detection of \ion{Si}{iii} (not included in Table\,\ref{tab:lines} due to heavy saturation), and \ion{Si}{iv} strongly suggests the presence of gas that is partially or predominantly ionised.  

While the metal absorption shows no change in the ionisation structure in the gas on the lines of sight \seven\ to \one, the lines of sight to \eleven\ and \new\ distinguish themselves from the other sight-lines. 
Every line of sight samples a mix of CNM, WNM, and WIM, of which the fraction of ionised gas appears to be higher in the lines of sight \seven\ to \one. 

It would have been favourable if the \ion{H}{i} column densities would
have been known from 21 cm data. Such high spatial resolution data do
not exist. The Galactic All Sky Survey, GASS \citep{McClure}, has a resolution
of only 16\arcmin, which does not help for our investigation.
ATCA interferometric 21\,cm data on the other hand are only available for
velocities $v > 30$ \kms.
Yet, the GASS 21\,cm profiles show over 30\arcmin\ a slight trend.
The high velocity resolution of the GASS 21\,cm data reveals a two-component profile at velocities of $+4$ and $-4$ \kms.
The intensity of the two components increases/decreases to opposite east/west directions overall in favour of a stronger \ion{H}{i} emission to lower RA.
These velocity components are not resolved in our FUSE, or STIS spectra.
Considering the changes in the N(\ion{O}{i}), however, it is likely that
our pencil beam spectra could be going through the cloud boundary,
where the column density summed over the two components is less to e.g.,
\new, compared to the other sight-lines.

\clearpage

\begin{landscape}
\begin{table}
\caption{Column densities $\log N$ ($N$ in cm$^{-2}$) of Galactic foreground gas, 1\,$\sigma$ errors, and Doppler parameter $b$\,[km\,s$^{-1}$] measured from the \cog.$^{\mathrm{a}}$} 
\label{tab:N} 
\centering
\begin{tabular}{l lll lll lll lll lll lll} 
\hline\hline 
        & \multicolumn{2}{c}{\eleven} & & \multicolumn{2}{c}{\new} & & 
\multicolumn{2}{c}{\seven} & & \multicolumn{2}{c}{\six} & & 
\multicolumn{2}{c}{\four} & & \multicolumn{2}{c}{\one} & \\ 
        & \multicolumn{2}{c}{(FUSE)} & & \multicolumn{2}{c}{(FUSE)} & & 
\multicolumn{2}{c}{(FUSE+STIS)} & & \multicolumn{2}{c}{(FUSE+STIS)} & & 
\multicolumn{2}{c}{(FUSE+STIS)} & & \multicolumn{2}{c}{(FUSE+STIS)} & \\ 
 
Species & $b$ & $\log N$ & & $b$ & $\log N$ & & $b$ & $\log N$ & & $b$ & $\log N$ & & $b$ & $\log N$ & & $b$ & $\log N$ & \\ 
\hline 
 & & & & & & & & & & & & & & & & & \\ 
H$_2$$^{\mathrm{F}}$ & & & & & & & & & & & & & & & & & \\ 
$J=0$  
& $4^{+1}_{-3}$ & $15.3^{+1.6}_{-0.1}$ &  
& $2^{+1}_{-1}$ & $15.4^{+1.1}_{-0.1}$ &  
& $3^{+4}_{-1}$ & $17.3^{+0.1}_{-1.7}$$^{\mathrm{c}}$ &  
& $3^{+8}_{-1}$ & $17.8^{+0.1}_{-2.1}$ &  
& $2^{+1}_{-1}$ & $17.9^{+0.2}_{-0.2}$$^{\mathrm{c}}$ &  
& $4^{+9}_{-1}$ & $17.9^{+0.1}_{-2.0}$ \\ 
$J=1$  
& $6^{+1}_{-4}$ & $15.6^{+1.8}_{-0.1}$ &  
& $4^{+1}_{-2}$ & $15.6^{+1.3}_{-0.2}$ &  
& $3^{+3}_{-1}$ & $17.3^{+0.3}_{-1.5}$ &  
& $3^{+1}_{-1}$ & $17.5^{+0.1}_{-0.1}$ &  
& $2^{+9}_{-1}$ & $17.7^{+0.1}_{-2.3}$ &  
& $2^{+1}_{-1}$ & $17.7^{+0.1}_{-0.2}$ \\ 
$J=2$  
& $4^{+1}_{-3}$ & $15.0^{+1.9}_{-0.1}$ &  
& $5^{+3}_{-2}$ & $14.5^{+0.2}_{-0.1}$ &  
& $4^{+2}_{-3}$ & $15.0^{+1.9}_{-0.2}$ &  
& $4^{+1}_{-2}$ & $15.5^{+1.5}_{-0.3}$ &  
& $5^{+1}_{-4}$ & $15.0^{+2.1}_{-0.1}$ &  
& $4^{+1}_{-3}$ & $15.1^{+2.0}_{-0.1}$ \\ 
$J=3$  
& $8^{+6}_{-2}$ & $14.5^{+0.1}_{-0.1}$ &  
& $6^{+4}_{-3}$ & $14.3^{+0.2}_{-0.1}$ &  
& $5^{+9}_{-3}$ & $14.5^{+0.6}_{-0.1}$  &  
& $4^{+1}_{-3}$ & $15.1^{+1.9}_{-0.2}$ &  
& $4^{+1}_{-1}$ & $14.9^{+0.1}_{-0.1}$ &  
& $5^{+4}_{-1}$ & $14.8^{+0.2}_{-0.1}$ \\ 
$J=4$  
& $>2$ & $<13.6$ &  
& $>2$ & $<13.6$ &  
& $>2$ & $<13.9$ &  
& $5^{+9}_{-3}$ & $14.1^{+0.1}_{-0.1}$ &  
& $4^{+1}_{-2}$ & $13.8^{+0.2}_{-0.1}$ &  
& $2^{+12}_{-1}$ & $13.8^{+0.2}_{-0.1}$ \\ 
Total  
&  & $15.9^{+1.7}_{-0.1}$ &  
&  & $15.8^{+1.2}_{-0.1}$ &  
&  & $17.6^{+0.3}_{-1.5}$ &  
&  & $18.0^{+0.2}_{-0.6}$ &  
&  & $18.1^{+0.2}_{-0.4}$ &  
&  & $18.1^{+0.1}_{-0.6}$ \\ 
\hline 
 & & & & & & & & & & & & & & & & & \\ 
C\,{\sc i}$^{\mathrm{FS}}$  
& $3^{+1}_{-1}$ & $14.0^{+0.1}_{-0.1}$ &  
& $3^{+1}_{-1}$ & $13.8^{+0.1}_{-0.1}$ &  
& $3^{+1}_{-1}$ & $13.8^{+0.3}_{-0.1}$ &  
& $3^{+1}_{-1}$ & $13.9^{+0.3}_{-0.1}$ &  
& $3^{+1}_{-1}$ & $14.0^{+0.1}_{-0.1}$ &  
& $3^{+1}_{-1}$ & $14.2^{+0.1}_{-0.1}$ \\ 
C\,{\sc i*}$^{\mathrm{FS}}$  
&  & $<13.8$ &  
&  & $<13.8$ &  
&  & $13.5^{+0.2}_{-0.1}$ &  
&  & $13.5^{+0.3}_{-0.1}$ &  
&  & $13.6^{+0.2}_{-0.1}$ &  
&  & $13.6^{+0.2}_{-0.1}$ \\ 
C\,{\sc i**}$^{\mathrm{FS}}$  
&  & $<13.8$ &  
&  & $<12.9$ &  
&  & $<12.8$ &  
&  & $<12.9$ &  
&  & $<12.8$ &  
&  & $<12.8$ \\ 
O\,{\sc i}$^{\mathrm{FS}}$  
& $7^{+1}_{-1}$ & $17.3^{+0.4}_{-0.1}$ &  
& $7^{+1}_{-1}$ & $16.9^{+0.2}_{-0.1}$ &  
& $7^{+1}_{-1}$ & $17.6^{+0.1}_{-0.1}$ &  
& $7^{+1}_{-1}$ & $17.7^{+0.1}_{-0.5}$ &  
& $7^{+1}_{-1}$ & $17.6^{+0.2}_{-0.3}$ &  
& $7^{+1}_{-1}$ & $17.7^{+0.3}_{-0.3}$ \\ 
N\,{\sc i}$^{\mathrm{FS}}$  
&  & $16.4^{+0.1}_{-0.1}$ &  
&  & $16.2^{+0.4}_{-0.1}$ &  
&  & $16.8^{+0.1}_{-0.7}$ &  
&  & $16.3^{+0.4}_{-0.4}$ &  
&  & $16.4^{+0.1}_{-0.3}$ &  
&  & $16.8^{+0.4}_{-0.7}$ \\ 
Ar\,{\sc i}$^{\mathrm{F}}$  
&  & $14.8^{+0.6}_{-0.1}$ &  
&  & $14.3^{+0.2}_{-0.1}$ &  
&  & $14.8^{+0.8}_{-0.3}$ &  
&  & $14.9^{+0.8}_{-0.3}$ &  
&  & $14.8^{+1.0}_{-0.3}$ &  
&  & $14.8^{+0.9}_{-0.3}$ \\ 
\hline 
\\ 
\ion{Fe}{ii}$^{\mathrm{F}}$  
& $8^{+1}_{-1}$ & $14.8^{+0.1}_{-0.1}$ &  
& $7^{+1}_{-1}$  & $14.6^{+0.1}_{-0.1}$ &  
& $8^{+1}_{-1}$ & $14.8^{+0.1}_{-0.1}$  &  
& $8^{+1}_{-1}$ & $14.8^{+0.1}_{-0.1}$ &  
& $9^{+1}_{-1}$ & $14.8^{+0.1}_{-0.1}$ &  
& $8^{+1}_{-1}$ & $14.8^{+0.1}_{-0.1}$ \\ 
Mg\,{\sc ii}$^{\mathrm{S}}$  
&  &  &  
&  &  &  
&  & $15.7^{+0.1}_{-0.1}$ &  
&  & $15.9^{+0.1}_{-0.1}$ &  
&  & $15.8^{+0.1}_{-0.1}$ &  
&  & $15.8^{+0.1}_{-0.1}$ \\ 
Al\,{\sc ii}$^{\mathrm{FS}}$  
&  & $<14.5$ &  
&  & $<14.8$ &  
&  & $15.0^{+0.2}_{-0.9}$ &  
&  & $15.0^{+0.3}_{-0.9}$ &  
&  & $14.2^{+0.8}_{-0.7}$ &  
&  & $14.8^{+0.3}_{-0.8}$ \\ 
Si\,{\sc ii}$^{\mathrm{FS}}$ 
&  & $>15.1$$^{\mathrm{b}}$ &  
&  & $>14.9$$^{\mathrm{b}}$ &  
&  & $16.5^{+0.3}_{-0.6}$$^{\mathrm{c}}$ &  
&  & $16.5^{+0.3}_{-0.6}$$^{\mathrm{c}}$ &  
&  & $16.3^{+0.2}_{-0.6}$ &  
&  & $16.4^{+0.1}_{-0.7}$$^{\mathrm{c}}$ \\ 
Si\,{\sc iv}$^{\mathrm{S}}$ 
&  &  &  
&  &  &  
&  & $13.0^{+0.1}_{-0.1}$ &  
&  & $13.1^{+0.1}_{-0.1}$ &  
&  & $13.1^{+0.1}_{-0.1}$ &  
&  & $13.3^{+0.1}_{-0.2}$ \\ 
P\,{\sc ii}$^{\mathrm{FS}}$  
&  & $13.8^{+0.2}_{-0.2}$ &  
&  & $13.5^{+0.1}_{-0.2}$ &  
&  & $13.8^{+0.2}_{-0.1}$ &  
&  & $13.8^{+0.2}_{-0.2}$ &  
&  & $13.6^{+0.2}_{-0.1}$ &  
&  & $13.7^{+0.3}_{-0.1}$ \\ 
S\,{\sc ii}$^{\mathrm{S}}$  
&  &  &  
&  &  &  
&  & $15.5^{+0.4}_{-0.2}$ &  
&  & $15.5^{+0.5}_{-0.1}$ &  
&  & $15.4^{+0.4}_{-0.1}$ &  
&  & $15.5^{+0.5}_{-0.1}$ \\ 
S\,{\sc iii}$^{\mathrm{FS}}$ 
&  & $14.4^{+0.2}_{-0.1}$$^{\mathrm{b}}$ &  
&  & $14.4^{+0.2}_{-0.2}$$^{\mathrm{b}}$ &  
&  & $14.7^{+0.2}_{-0.1}$ &  
&  & $14.6^{+0.2}_{-0.1}$ &  
&  & $14.5^{+0.1}_{-0.1}$ &  
&  & $14.7^{+0.2}_{-0.1}$ \\ 
Mn\,{\sc ii}$^{\mathrm{S}}$ &  &  & &  &  &  
&  & $13.0^{+0.1}_{-0.1}$$^{\mathrm{b}}$ &  
&  & $13.0^{+0.1}_{-0.1}$$^{\mathrm{b}}$ &  
&  & $13.0^{+0.2}_{-0.1}$$^{\mathrm{b}}$ &  
&  & $13.1^{+0.1}_{-0.1}$$^{\mathrm{b}}$ \\ 
Ni\,{\sc ii}$^{\mathrm{S}}$ &  &  & &  &  &  
&  & $13.5^{+0.1}_{-0.1}$ &  
&  & $13.6^{+0.1}_{-0.1}$ &  
&  & $13.5^{+0.1}_{-0.1}$ &  
&  & $13.7^{+0.1}_{-0.1}$ \\ 
\hline 
\end{tabular} 
\begin{list}{}{} 
\item[$^{\mathrm{a}}$] Most of the metal column densities are based on the adopted $b$ from \ion{Fe}{ii} in each line of sight. \ion{N}{i}, and \ion{Ar}{i} are derived from the \cog\ of \ion{O}{i}, and the fine-structure levels of carbon from the \cog\ of \ion{C}{i}.
\item[$^{\mathrm{b}}$] Based on one absorption line 
\item[$^{\mathrm{c}}$] The best fit includes outliers (less than 32$\%$ of the data points). For the H$_2$ data, these outliers are all the absorptions from FUSE:SiC (see Sect.\,4.1). For the \ion{Si}{ii} data, the outlier is the weak line at 1020.7 {\AA} (see Sect.\,5).  
\item[$^{\mathrm{F}}$] Based on FUSE data only 
\item[$^{\mathrm{S}}$] Based on STIS data only 
\item[$^{\mathrm{FS}}$] Based on both FUSE and STIS for \seven\ to \one\ 
\end{list} 
\end{table} 
\end{landscape}
\addtocounter{table}{1} 
\section{Gas densities} 
\label{sect:Density} 

The available data allow us to derive the gas density in two ways. One possibility is based on the H$_2$/O abundance and models for the self-shielding of the gas. The other uses the level of collisional excitation of C\,{\sc i} as derived from the absorption lines. In the latter case the result is not only limited by the accuracy of column densities of the fine-structure levels of neutral carbon, but also by the derived $T_{\rm exc}$ from H$_2$. 
 
\subsection{Density variations derived from H$_2$} 
\label{subsect:DensityH2} 
 
The observed column density variations along our sight-lines observed for the 
various ions and H$_2$ suggest a change in the physical conditions in the 
foreground gas at relatively small spatial scales. For the metal ions, the 
observed column densities depend on three physical parameters: the volume 
density 
of the respective element in the gas, the fractional abundance of the observed 
ionisation state of that element (reflecting the ionisation conditions in the 
gas), and the thickness of the absorbing gas layer.  
These three quantities may vary between adjacent sight-lines, so that the ion 
column densities and their variations alone tells us very little about the 
actual (total) particle-density variations $\Delta n_{\rm H}$ in the gas. Yet it is these density variations that we wish to know in order to study the density structure of the ISM at small scales. 
 
As we show below, it is possible to reconstruct the particle density distribution in the gas by {\it combining} our ion column-density measurements with the H$_2$ measurements in our data sample, and assuming that the molecular abundance is governed by an H$_2$ formation-dissociation equilibrium (FDE). 
In FDE, the neutral to molecular hydrogen column density ratio is given by 
 
\begin{equation} 
\frac{N({\rm H\,I})}{N({\rm H}_2)} = 
\frac{\langle k \rangle \,\beta}{R\,n_{\rm H}} \ \ \ , 
\end{equation} 
 
\noindent 
where $\langle k \rangle \approx  0.11$ is the probability that the molecule is 
dissociated after photo absorption, $\beta$ is the photo-absorption rate per 
second within the cloud, and $R$ is the H$_2$ formation coefficient on dust grains in units cm$^{3}$\,s$^{-1}$.  
 
Unfortunately, the local H\,{\sc i} column density cannot be obtained directly 
from our absorption 
measurements because the damped H\,{\sc i} Ly\,$\beta$ absorption in our FUSE data represents a blended composite of all absorption components in the Milky Way and LMC along the line of sight. 
However, as mentioned in the previous section, neutral oxygen has the same 
ionisation potential as neutral hydrogen, and both atoms are coupled by a strong 
charge-exchange reaction in the neutral ISM, so that O\,{\sc i} serves as an 
ideal tracer for H\,{\sc i}. 
If we now solve the previous equation for 
$n_{\rm H}$ and assume that $N$(H\,{\sc i}$)=z_{\rm O}\, 
N$(O\,{\sc i}) with $z_{\rm O}$ as the local hydrogen-to-oxygen 
ratio, we obtain 
 
\begin{equation} 
n_{\rm H} = \frac{N({\rm H}_2)}{N({\rm O\,I})}\, 
\frac{\langle k \rangle \,\beta}{z_{\rm O}\,R} \ \ \ . 
\end{equation} 
 
For interstellar clouds that are optically thick in H$_2$ (i.e., log 
$N$(H$_2)\gg 14$) H$_2$ line self-shielding has to be taken into account. The 
self-shielding reduces the photo absorption rate in the cloud interior and 
depends on the total H$_2$ column density in the cloud. \citet{Draine:Bertoldi} find that 
the H$_2$ self-shielding can be characterised by the relation 
$\beta=S\,\beta_0$, where $S=(N_{\rm H_2}/10^{14}$cm$^{-2})^{-0.75}<1$ is the 
self-shielding factor and $\beta_0$ is the photo absorption rate at the edge of 
the cloud (which is directly related to the intensity of the ambient UV 
radiation field). 
 
Because all our sight-lines are passing with small separation through the same local interstellar gas 
cloud, the 
respective parameters $\beta_0$, $R$, and $z_{\rm O}$ should be identical along 
these sight-lines. The observed quantities $N$(H\,{\sc i}) and $N$(O\,{\sc i}) 
instead vary from sight-line to  
sight-line due to the local density variations $\Delta n_{\rm H}$ and different 
self-shielding factors in the gas. 
If we now consider Eq.\,(2) and assume $\beta=S(N($H$_2))\,\beta_0$,  
for two adjacent lines of sight (LOS1 and LOS2) through the same cloud, we 
obtain for the local {\it density ratio} $\xi$ between LOS1 and LOS2 
 
\begin{equation} 
\xi = \frac{n_{\rm H,LOS1}}{n_{\rm H,LOS2}} = 
\frac{N({\rm \ion{O}{i}})_{\rm LOS2}}{N({\rm \ion{O}{i}})_{\rm LOS1}}\, 
\left(\frac{ N({\rm H}_2)_{\rm LOS1}}{N({\rm H}_2)_{\rm LOS2}} 
\right)^{0.25} . 
\end{equation} 
 
A small rise in $n_{\rm H}$ only slightly increases the H$_2$ grain formation 
$Rn_{\rm H}$, but it substantially reduces the photo 
dissociation rate $\langle k \rangle \beta_0 S$ due to the increased H$_2$ 
self-shielding, so that in the end 
$\Delta N$(H$_2) \propto \Delta n_{\rm H}$$^4$ while  
$\Delta N$(O\,{\sc i}$) \propto n_{\rm H}$. 
Therefore, the H$_2$/O\,{\sc i} column density ratio represents a very sensitive 
tracer for density variations in the gas.  
 
\begin{figure}[tbp] 
\centering 
\includegraphics[scale=0.43]{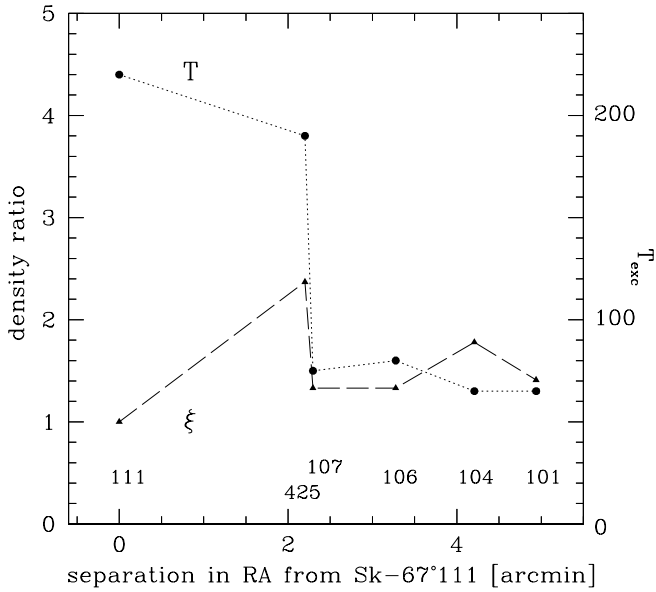} 
\caption{Density ratio, $\xi$, and the excitation temperature, $T_{\rm exc}$, with angular separation from \eleven\ (for separation in pc see Fig.\,\ref{fig:3454f3}). The parameter $\xi$ (filled triangles) is calculated according to Eq.\,(3) based on \eleven\ (hence $\xi=1$ at separation 0). $T_{\rm exc}$ (filled circles) is taken from Fig.\,\ref{fig:3454f5}. The sight-lines are marked with the three last numbers of the star name.  
While $T_{\rm exc}$ suggests the existence of a confined core with molecular hydrogen in the sight-lines \one\ to \seven, $\xi$ appears to be significantly higher towards \new.} 
\label{fig:3454f8} 
\end{figure}

We now use Eq.\,(3) to study the density variations in the local gas 
towards our six lines of sight. 
In Fig.\,\ref{fig:3454f8} we plot the relative density to 
\eleven\ together with the derived $T_{\rm exc}$ versus the separations in RA coordinates.  
While the excitation temperature suggests a cold core in the sight-lines \one, \four, \six, and \seven, the total mean density appears to be lower by a factor of almost 2 compared to \new.  
 
While the column density variations of different excitational levels of H$_2$ suggest that we are observing a high density core of the H$_2$ between the sight-lines \one\ to \six, the result from the relative abundances together with the derived total density variation point to a higher mean density to \new. 
 
This might be due to a smaller total absorber pathlength in the direction of \new, which can include a slightly more confined region, for which the mean density averaged over the pathlength is higher compared to the other sight-lines, or due to the higher fraction of neutral gas towards \new, on which we base $\xi$. 
 
If we analyse the density $n_{\rm H}$ directly from Eq.\,(3) by using the typical values observed for $R=3\times 10^{-17} \,\rm {cm^{3}}\,\rm {s^{-1}}$ and $\beta_0=5\times 10^{-10} \,\rm {s^{-1}}$ \citep{S78}, and the total neutral hydrogen column density $N(\ion{H}{i})$ derived from $N(\ion{O}{i})$ assuming solar oxygen abundance, we obtain $n_{\rm H}\simeq 2 \,\rm {cm^{-3}}$ on average. 
This represents the lower limit required for the H$_2$ to exist in the FDE. The observed $N(\ion{O}{i})$ represents the indirect measure of the total $N(\ion{H}{i})$ along a line of sight. It is likely that only a fraction of this \ion{H}{i} is available for the H$_2$ formation, and that the H$_2$ self-shielding is less efficient than assumed because of a complex geometry of the H$_2$ structures. 

\subsection{Density from C\,{\sc i} excitation} 
 
The ground electronic state of C\,{\sc i} is split into three fine-structure levels. The two upper levels C\,{\sc i*} and C\,{\sc i**} are populated through collisional excitation and some UV-pumping \citep{deBoerCI}. 
We can therefore use the fine-structure levels of neutral carbon together with the excitation temperature derived from H$_2$, to derive the density $n_{\rm H}$ of the gas. 
 
Using the population of the fine-structure levels as a function of $n_{\rm H}$ and $T$, calculated by \cite{deBoerCI}, we find the densities in the range of $\sim 40-60$ cm$^{-3}$ on the lines of sight \one\ to \six, and a higher density of $\sim 80$ cm$^{-3}$ towards \seven\ (see Table\,\ref{tab:summary}). 
The densities derived in this way are dependent on the accuracy in the temperature, as well as the C\,{\sc i} column densities.  
For \new\ and \eleven\ we are only able to derive an upper limit for C\,{\sc i*}, and hence upper limits of $<126$ and $<65$ cm$^{-3}$, respectievly for $n_{\rm H}$ in these sight-lines.  
Note that these upper limits are not comparable. While the upper limit of C\,{\sc i*} towards \new\ was based on an absorption with $f$-value similar to the ones used towards \seven\ to \one, the C\,{\sc i*} upper limit for the \eleven\ sight-line was based on a weaker line, since the same spectral region was occupied by an absorption of a different velocity component. 
 
The column densities of C\,{\sc i} also allow us to calculate the ionisation 
balance, assuming that the abundance ratio of C to O is solar. One thus has 
\begin{equation} 
\frac{N(\ion{C}{ii})}{N(\ion{C}{i})} 
 = \frac{\Gamma}{\alpha(T)\ n_e} \ \ \ , \end{equation}
and with $N$(C)=(C/O)$_{\odot}\cdot N$(O) one obtains \begin{equation} 
\frac{N(\ion{C}{i})}{N(\ion{O}{i})} 
 = \frac{{\rm (C/O)_{\odot}}\ \alpha(T)\ n_e}{\Gamma} \ \ \ . 
\end{equation} 
Using (C/O)$_{\odot}\simeq$0.6 \citep{Asplund}, $\alpha(T\simeq 70\,\rm {K}) \simeq 12 \cdot 10^{-12}$  
cm$^3$\,s$^{-1}$ \citep{Pequignot} and $\Gamma({\rm C})=310 \cdot 10^{-12}$ s$^{-1}$ 
\citep{deBoer73}, one arrives at  
\begin{equation} 
\frac{N(\ion{C}{i})}{N(\ion{O}{i})} = 0.025\ n_e \ \ \ . 
\end{equation} 
Note that for $T\simeq 210$\,K $\alpha$ would be a factor 1.7 smaller. 
And for strong shielding of UV radiation, $\Gamma$ would be up to a 
factor of 2 smaller (see \cite{deBoer73}). The column density ratios of 
C\,{\sc i} to O\,{\sc i} lead to values of $n_e$ as given in Table\,\ref{tab:summary}.  
We used the value of $\alpha$ for $T=70$\,K for the sight-lines \seven\ to \one\ and for the other two sight-lines the 1.7 smaller $\alpha$. 
 
These values, in combination with $n$(C\,{\sc i}), indicate a
larger ionisation fraction towards \eleven\ and \new.

\section{Conclusions} 
 
\label{sect:summary} 
We have presented UV absorption line measurements of Galactic H$_2$, \ion{C}{i}, \ion{N}{i}, O\,{\sc i}, Al\,{\sc ii}, Si\,{\sc ii}, P\,{\sc ii}, S\,{\sc iii}, Ar\,{\sc i}, and \ion{Fe}{ii} towards the six LMC stars \eleven, \new, \seven, \six, \four, and \one, and analysed the properties of the Galactic disc 
gas in these lines of sight. Our sight-lines were chosen within $5\arcmin$ in RA, with an almost constant Dec, to allow us to investigate the small-scale structures of the interstellar gas within $<1.5$\,pc (assuming that the foreground gas exists at a distance of $<1$\,kpc), with the smallest separation corresponding to $<0.08$\,pc. 
For four of the sight-lines, \one, \four, \six, and \seven, we also analysed STIS spectral data, from which we have further determined the column densities of \ion{C}{i*}, \ion{C}{i**}, Mg\,{\sc ii}, Si\,{\sc iv}, S\,{\sc ii}, Mn\,{\sc ii}, and Ni\,{\sc ii}. 
The STIS data provide additional absorption lines for some of the species already available in the FUSE spectra, allowing us to have a more reliable determination of the column densities. 
 
In the spectral range of FUSE we are able to gain information from the different excitation levels of H$_2$, which provides us with valuable information about the fine structure of the gas, because H$_2$ arises in the coolest regions of the interstellar clouds, and enables us to derive the physical properties of the gas. 
The H$_2$ absorptions show considerable variation between our sight-lines. The lines of sight to \one\ and \four\ contain a significantly higher amount of H$_2$, about 2\,dex in their column densities, compared to \eleven.  
The rotational excitation of H$_2$ towards \one\ to \seven\ agrees with a core-envelope structure of the H$_2$ gas with an average temperature of $T_{\rm exc}\sim70$\,K, typical for CNM. This, and the low Doppler parameter of the lower $J$-levels, indicates that the molecular hydrogen probably arises in confined dense regions in these sight-lines, where it is self-shielded from UV-radiation. The gas to \new\ and \eleven, on the other hand, appears to be fully thermalised, with $T_{\rm exc}\sim200$\,K. 
This could suggest that within the $<5\arcmin$, we sample a core of a molecular hydrogen cloud, and reach the edge between \seven\ and \eleven.  
After including the \new\ sight-line in our study, however, the spatial variation is not a smooth change from \one\ to \eleven\ (Fig.\,\ref{fig:3454f8}), and density fluctuation on scales smaller than the extent of 5\arcmin\ cannot be excluded.  
 
In order to trace these variations back to the actual density variations and changes in the physical properties of the gas we have derived the densities partly based on the formation-dissociation equilibrium of molecular hydrogen using the measured H$_2$ and O\,{\sc i} column densities, and partly based on the excitation of the fine-structure levels of neutral carbon. For the latter we have also used the derived $T_{\rm exc}$(H$_2$), assuming that C\,{\sc i} and H$_2$ exist in the same physical region. 
The density derived in this way is $n_{\rm H} \simeq 40-80 \,\rm cm^{-3}$ towards \one\ to \seven, for which we have relatively accurate measurements for column densities of C\,{\sc i} and C\,{\sc i*}, as well as $T_{\rm exc}$. This is well above the value $n_{\rm H} \simeq 2 \,\rm cm^{-3}$ derived based on a formation-dissociation equilibrium of H$_2$. This is most likely because the available hydrogen for molecular formation is less than the total Galactic $N(\ion{H}{i})$ derived from N(\ion{O}{i}) along a line of sight.  
Moreover, the H$_2$ self-shielding could be overestimated (and thus $n_{\rm H}$ underestimated) because of a complex absorber geometry. 
Thus that value for $n_{\rm H}$ gives a minimum density required for the existence of the H$_2$.  
 
On all sight-lines $N(\rm {H_2}) \ll N(\ion{H}{i})$ . If we assume 
that $[N_{\rm H}]_{\rm tot}$ is proportional to $n_{\rm H}$ (from \ion{C}{i}) in the same way 
as for the inner region of a molecular cloud (meaning 
$[n_{\rm H}/N(\ion{H}{i})]_{\rm tot}=[n_{\rm H}/N(\ion{H}{i})]_{\rm mol}$), 
we can derive from Eq.\,(2) the column density of \ion{H}{i} that coexists with H$_2$ in FDE. 
We so find that this is 18\% of the total $N(\ion{H}{i})$ (averaged over the sight-lines \one\ to \seven). 
It is of course likely that the partly molecular core is more dense because it is confined, and $n_{\rm H}$ not linearly related to the $N(\ion{H}{i})$ as assumed above. 
Furthermore, $[n_{\rm H}]_{\rm tot}$ as derived from \ion{C}{i} is the averaged density of the molecular part of the gas, where \ion{C}{i} exists, and again not related to all the \ion{H}{i} gas along a line of sight as assumed above. 
Thus, the $N(\ion{H}{i})_{\rm tot}$ together with the $n_{\rm H}=40-80\,\rm cm^{-3}$ give only an upper limit for the pathlength of the partly molecular cloud to be $D_{\rm {mol}}=0.5-1.8$\,pc (included in Table\,\ref{tab:summary}). 
These derived pathlengths are of the same size as (or smaller than) our lateral extent of the sight-lines ($<1.5$\,pc). 
This derived upper limit for $D_{\rm {mol}}$ also agrees with the higher density ratio for the \new\ sight-line, derived in Sect.\,\ref{sect:Density}.1, which otherwise is difficult to explain if we consider this sight-line to sample the edge of the same cloud with the core on the \one\ to \seven\ sight-lines. Based on $\xi$ and an average density of $n_{\rm H}=60\,\rm cm^{-3}$, $D_{\rm {mol}}$ in the direction of \new\ and \eleven\ is estimated to be 0.1 and 0.6\,pc, respectively. Given these small sizes, the H$_2$ patches observed on these six lines of sight are not necessarily connected. The small pathlength is further important for the shielding of the molecular gas, given the small amount of \ion{H}{i} that is possibly available in FDE with H$_2$. 
 
We thus conclude that the H$_2$ observed along these six lines of sight exists in rather small cloudlets with an upper size $D_{\rm {mol}}=0.5-1.8$, and possibly even less, $<0.1$\,pc, as implied by the \new\ sight-line. 
Part of the absorbing gas is possibly located at a distance closer than the 1\,kpc assumed thus far, resulting in an even smaller spatial scale. This agrees with the sub-pc structure of molecular clouds previously detected by e.g., \cite{Pan}, \cite{Lauroesch:Meyer2000}, \cite{Richter:2003b,Richter:2003a}, and \cite{Marggraf}. 
On a smaller scale of 5 to 104 AU however, \cite{Boisse2009} found no variation in the column density of the H$_2$, observed over a period of almost five years towards the Galactic high-velocity star HD\,34078. This despite the variations in the CH column densities found over the same period. They concluded that while the variation in CH and CH$^+$ is due to the chemical structure of a gas that is in interaction with the star, the H$_2$, located in a quiescent gas unaffected by the star, remains homogeneously distributed. Hence no small-scale density structure was proposed in the quiescent gas within this scale.
Our spatial resolution limits us to the smallest separation of 17\arcsec. We have therefore no further information about the AU-scale structure of the molecular gas in these lines of sight.

The absorber pathlength through the disc has previously been found to be divided into smaller cloudlets of \ion{H}{i}-absorbers \citep{Welty}, which are not necessarily physically connected to the cloud with the observed H$_2$. 
While the clumpy nature of H$_2$ is verified by different studies, the Galactic \ion{H}{i} gas, when sampled over a line of sight, normally does not show variations on small scales, since such effects statistically cancel out and smooth the total observed \ion{H}{i} (or \ion{O}{i} in our case). 
If such cloudlets exist in the Galactic disc component, their LSR velocity should be similar, and therefore would not be resolved in our spectra. 
Our derived density ratio $\xi$, as well as the relative abundances, suggest that the number of these absorbers might differ along the sight-lines, being less towards \new\ and \eleven. 
This interpretation also agrees with the higher electron density derived for these lines of sight. 

We summarised in Sect.\,\ref{sect:Introduction} the detection of small-scale structure by different independent methods. 
This paper presents yet another approach to identify the structure of the gas on small scales. 
FUV and UV spectral data, with the large range of transitions particularly of H$_2$ and \ion{O}{i} together with the fine-structure of \ion{C}{i}, provide a sensitive approach to study the physical properties of the gas on small scales. 
However, our spatial resolution is limited to the separation of our background sources, and does not allow us to directly detect the fine structure that might exist on even smaller scales. 
Our findings are consistent with the small-scale structure found earlier in the interstellar medium, and suggest inhomogeneity in the interstellar Galactic gas down to, and possibly below, scales of 0.1\,pc. 

\begin{acknowledgements} 
We thank Ole Marggraf for his careful proofreading of this paper.
SNS was supported through grant BO779/30 by the German Research Foundation (DFG).
\end{acknowledgements}

\bibliographystyle{aa}
\bibliography{bib3454}

\longtab{4}{
\begin{longtable}{lrlll} 
\caption{\label{tab:lines} List of absorption lines, used for the column density determination in one or more sight-lines.}\footnotemark[1] \\
\hline\hline 
Species & $\lambda_0$[\AA]\footnotemark[2] & $f (\times 10^{-3})$\footnotemark[2] & Instrument\footnotemark[3] & detector \\ 
\hline 
\endfirsthead
\caption{continued.}\\
\hline\hline
Species & $\lambda_0$[\AA]\footnotemark[2] & $f (\times 10^{-3})$\footnotemark[2]  & Instrument\footnotemark[3] & detector \\ 
\hline
\endhead
\hline
\endfoot
\ion{C}{i} 
 & 945.19 & 152.223  & FUSE & SiC \\
 & 1139.79 & 12.3 & FUSE & LiF \\ 
 & 1157.91 & 21.2  & FUSE & LiF \\ 
 & 1158.32 & 6.55   & FUSE & LiF \\ 
 & 1276.48 & 5.89   & STIS &  \\ 
 & 1280.14 & 26.3   & STIS &  \\ 
 & 1328.83 &  75.7756  & STIS &  \\ 
 & 1560.31  & 77.4086   & STIS &  \\ 
 & 1656.93 & 149.0  & STIS &  \\ 
 &  &  &  &  \\ 
\ion{C}{i*}
 & 1279.89 & 14.3  & STIS &  \\ 
 & 1560.68 & 58.1   & STIS &  \\ 
 & 1656.27 & 62.1   & STIS &  \\ 
 & 1657.38 & 37.1  & STIS &  \\ 
 & 1657.91 & 49.4   & STIS &  \\ 
 &  &  &  &  \\ 
\ion{C}{i**}
 & 1139.76 & 9.54   & FUSE & LiF \\ 
 & 1329.34 & 75.8   & STIS &  \\ 
 & 1657.18 & 149.0   & STIS &  \\ 
 &  &  &  &  \\ 
\ion{N}{i} 
 & 952.30 & 2.29000  & FUSE & SiC \\
 & 952.42 & 1.97000  & FUSE & SiC \\
 & 952.52  &  0.518   & FUSE & SiC \\
 & 953.42 & 12.9300  & FUSE & SiC \\
 & 953.66 & 24.6900  & FUSE & SiC \\
 & 953.97 & 33.0700  & FUSE & SiC \\
 & 954.10 & 4.00000  & FUSE & SiC \\
 & 963.99 & 12.4100  & FUSE & SiC \\
 & 1134.17 & 14.6000  & FUSE & LiF \\ 
 & 1134.41 & 28.6500  & FUSE & LiF \\ 
 & 1134.98 & 41.5900  & FUSE & LiF \\
 & 1199.55 & 131.820  & STIS &  \\ 
 &  &  &  &  \\ 
\ion{O}{i} 
 & 922.20 & 0.245  & FUSE & SiC \\ 
 & 924.95 & 1.54500  & FUSE & SiC  \\ 
 & 925.45 & 0.354  & FUSE & SiC \\ 
 & 930.26 & 0.537  & FUSE & SiC \\ 
 & 936.63 & 3.650  & FUSE & SiC \\ 
 & 971.74 & 11.6  & FUSE & SiC \\ 
 & 974.07 & 0.0156  & FUSE & SiC \\ 
 & 976.45 & 3.306  & FUSE & SiC \\ 
 & 1039.23 & 9.060  & FUSE & LiF \\ 
 & 1302.17 & 48.0100  & STIS & \\ 
 & 1355.60 & 0.00116  & STIS &  \\ 
 &  &  &  &  \\ 
\ion{Mg}{ii}
 & 1239.93 & 0.6320   & STIS &  \\ 
 & 1240.39 & 0.3560   & STIS &  \\ 
 &  &  &  &  \\ 
\ion{Al}{ii}
 & 935.27 &  2.51803  & FUSE & SiC \\ 
 & 1670.79  & 1738.12  & STIS &  \\ 
 &  &  &  &  \\ 
\ion{Si}{ii}
 & 1020.70  & 16.7920   & FUSE & LiF \\ 
 & 1190.42  & 291.945    & STIS &  \\
 & 1193.29  & 582.444    & STIS &  \\
 & 1304.37  & 86.4170    & STIS &  \\
 & 1526.71  & 132.814    & STIS &  \\
 &  &  &  &  \\ 
\ion{Si}{iv}
 & 1393.76  & 513.0   & STIS &  \\
 & 1402.77  & 254.0   & STIS &  \\
 &  &  &  &  \\ 
\ion{P}{ii}
 & 961.04 & 348.544  & FUSE & SiC \\
 & 963.80 & 1458.86 & FUSE & SiC \\
 & 1152.82 & 2.45041  & FUSE & LiF \\
 &  &  &  &  \\ 
\ion{S}{ii}
 & 1250.58 &  5.43112   & STIS &  \\
 & 1253.81 &  10.9086   & STIS &  \\
 & 1259.52 &  16.5881   & STIS &  \\
 &  &  &  &  \\ 
\ion{S}{iii}
 & 1012.49 &  43.8134   & FUSE & LiF \\
 & 1190.20 &  23.6254   & STIS &  \\
 &  &  &  &  \\ 
\ion{Ar}{i}
 & 1048.22  & 262.753   & FUSE & LiF \\
 & 1066.66  & 67.4488   & FUSE & LiF \\
 & & & & \\
\ion{Mn}{ii}
 & 1197.18  & 217.0   & STIS &  \\
 & & & & \\
\ion{Fe}{ii}
 & 1055.26     & 7.500   & FUSE & LiF \\
 & 1062.15 	 &   3.802   & FUSE & LiF \\
 & 1063.18   & 54.7513   & FUSE & LiF \\
 & 1081.87   & 12.5841   & FUSE & LiF \\
 & 1096.88   & 32.6469   & FUSE & LiF \\
 & 1112.05  & 4.46   & FUSE & LiF \\
 & 1121.97   & 28.9746   & FUSE & LiF \\
 & 1125.45   & 15.5838   & FUSE & LiF \\
 & 1127.10 	 &   2.8  & FUSE & LiF \\
 & 1133.67  & 4.72   & FUSE & LiF \\
 & 1142.37  & 4.01  & FUSE & LiF \\
 & 1143.23  &  19.2251  & FUSE & LiF \\
 & 1144.94  &  83.0268  & FUSE & LiF \\
 & 1608.45  &  57.7553  & STIS &  \\
 & & & & \\
\ion{Ni}{ii}
 & 1370.13 &  76.9551  & STIS &  \\
 & 1454.84 &  32.2986  & STIS &  \\

\addtocounter{footnote}{-3}
\stepcounter{footnote}\footnotetext[1]{Due to the varying S/N for different data and the differences in the shape of the continua, a few of the listed absorptions are not used in all six sight-lines.}\footnotetext[2]{The rest wavelength $\lambda_0$, and $f$-values are from \citet{Morton}.}\footnotetext[3]{No STIS data are available for \eleven\ and \new.}
\end{longtable}
}

\begin{table*} 
\caption{Physical conditions in the Milky Way foreground gas 
as derived from the data.} 
\label{tab:summary} 
\begin{tabular}{l l l l l l l l} 
\hline \hline
star & \eleven\ & \new\ & \seven\ & \six\ & \four\ & \one\ & comments\\ 
\hline 
$\log N(\ion{H}{i})$ & 20.6 & 20.2 & 20.9 & 21.0 & 20.9 & 21.0 & Sect.\,6.1$^{\mathrm{a}}$ \\  
$\log N \rm (H_2)$ & 15.9 & 15.8 & 17.6 & 18.0 & 18.1 & 18.1 & Sect.\,4\\  
 
$T_{\rm exc}$ (H$_2$) [K] & 220 & 190 & 75 & 80 & 65 & 65 & Sect.\,4.2\\  
$\xi$ & 1.0 & 2.4 & 1.3 & 1.3 & 1.8 & 1.4 & Sect.\,6.1 \\  
$n_{\rm H}$(C\,{\sc i}) [cm$^{-3}$] & $<$65 & $<$125 & 80 & 55 & 65 & 40 & Sect.\,6.2 \\ 
$n_e$ [cm$^{-3}$] & 0.045 & 0.072 & 0.007 & 0.007 & 0.005 & 0.014 & Sect.\,6.2\\ 
$D_{\rm tot}$ [pc] & & & 3.5 & 6.5 & 4.5 & 9 & Sect.\,7\\ 
$D_{\rm mol}$ [pc] & 0.6$^{\mathrm{b}}$ & 0.1$^{\mathrm{b}}$ & 0.5 & 1.1 & 0.8 & 1.8 & Sect.\,7\\ 
\hline 
Shielding as derived from H$_2$:\\ 
 & low & low &  &  &  &  & Sect.\,4.2\\ 
\hline 
Depletion\\ 
Fe,Mn,Ni &  &  & high & high & high & high & Sect.\,5\\ 
\hline 
\end{tabular} 
\begin{list}{}{} 
\item[$^{\mathrm{a}}$] Derived from \ion{O}{i} and (O/H)$_{\odot}=-3.34$ \citep{Asplund}. 
\item[$^{\mathrm{b}}$] Based on $\xi$ and a reference LOS averaged over the values for \one\ to \seven. 
\end{list} 
\end{table*} 

\end{document}